\documentstyle[aps,preprint,eqsecnum,epsf]{revtex}
\input psfig.sty

\begin{document}
\draft

\title{Performance of Discrete Heat Engines and Heat Pumps in Finite Time.}

\author{Tova Feldmann and Ronnie Kosloff}

\address{
Department of Physical Chemistry 
the Hebrew University, Jerusalem 91904, Israel\\
}

\maketitle

\begin{abstract}
The performance in finite time of a discrete heat engine with  internal friction
is analyzed. The working fluid of the engine is composed of an 
ensemble of noninteracting  two level systems.
External work is applied by changing the external field and 
thus the internal energy levels.
The friction induces a minimal cycle time. The power output
of the engine is optimized with respect to time allocation between the 
contact time with the hot and cold baths as well as the adiabats.
The engine's performance is also optimized with respect to the external fields.
By reversing the cycle of operation a heat pump is constructed.
The performance of the engine as a heat pump is also optimized.
By varying the time allocation between the adiabats and the contact time
with the reservoir a universal behavior can be identified.
The optimal performance of the engine when the cold
bath is approaching absolute zero is studied.
It is found that the optimal cooling rate 
converges linearly to zero when the temperature approaches absolute zero.
\end{abstract}

\section{Introduction}
\label{sec:introduction}

Analysis of heat engines has been a major source of thermodynamic insight.
The second law of thermodynamics resulted from Carnot's study 
of the reversible heat engine \cite{carnot}.
Study of the endo-reversible Newtonian engine 
\cite{curzon75} began the field
of finite time thermodynamics \cite{salamon77,salamon80,andresen83,bejan96}.
Analysis of a virtual  heat engine by Szilard led to the connection
between thermodynamics and information theory 
\cite{szilard29,brillouin}.
Recently this connection has been extended to the regime of quantum
computation  \cite{lloyd}.  
 
Quantum models of heat engines show a remarkable similarity
to engines obeying macroscopic dynamics.
The Carnot efficiency is a well established limit for the efficiency
of lasers as well as other quantum engines 
\cite{geusic59,scovil59,geusic67,levine74,benshaul79}.
Moreover, even the irreversible operation of quantum engines with
finite power output has many similarities to macroscopic 
endo-reversible engines \cite{k24,geva0,geva1,geva2,wu98}.

It is this line of thought that serves as a motivation
for a detailed analysis of a discrete four stroke quantum engine.
In a previous study \cite{feldmann96}, the same  model served
to find the limits of the  finite time performance of such an engine but with
the emphasis on power optimization.
In that study the working medium was composed of  discrete level systems
with the dynamics governed by a master equation.
The purpose was to gain insight into the optimal
engine's performance with respect to time allocation when external 
parameters such as: the applied fields, the bath temperatures
and the relaxation rates were fixed.

The present analysis emphasizes the reverse operation of the heat engine
as a heat pump. For an adequate description of this mode of operation
inner friction has to be a consideration. Without it the model is deficient
with respect to optimizing the cooling power. Another addition is the 
optimization of the external fields. This is a common practice 
when cold temperatures are approached. With the addition of these 
two attributes,
the four stroke quantum model is analyzed both 
as a heat engine and as a refrigerator.

Inner friction is found to have a profound influence on performance 
of the refrigerator.
A direct  consequence of the friction is a lower bound on the cycle time.
This lower bound  excludes the non-realistic global optimization solutions
found for frictionless cases \cite{feldmann96}
where the cooling power can be optimized beyond bounds.
This observation, has led to the suggestion of replacing 
the optimization of the cooling power
by the optimization of the cooling efficiency 
per unit time \cite{Velasco,Velasco97,Velasco98,yan98}.
Including friction is therefore essential for more realistic models of 
heat engines and refrigerators with the natural optimization goal becomes 
either the power output or the cooling power.
The source of friction is not considered explicitly in the present model.
Physically friction is the result of non-adiabatic phenomena 
which are the result of the rapid change in the energy level 
structure of the system.
For example friction can be caused
by the missalignement of the external fields with 
the internal polarization of the working medium.  For a more explicit
description  of the friction the interactions between the individual 
particles composing the working fluid have to be  considered.
The present model is a microscopic 
analogue of the Ericsson refrigeration cycle
\cite{chen98} where the working fluid consists of magnetic salts.
The advantage of the  microscopic model is that the use of the 
phenomenological heat transfer laws can be avoided \cite{geva0}. 
The results of the present model are  compared to a recent analysis of
macroscopic chillers \cite{gordon97}. In that study, a universal modeling
was demonstrated. It is found that the discrete quantum version
of heat pumps has behavior similar  to that of macroscopic chillers.

There is a growing interest  in the topic of cooling
atoms and molecules to temperatures very close 
to absolute zero \cite{cohen}. Most of the analysis of the
cooling schemes employed are based on quantum dynamical models.
New insight can be gained by employing a thermodynamic perspective.
In particular the temperatures achieved are so low that
the third law of thermodynamics has to be considered.
The discrete level heat pump can serve as a model to study
the third law limitations.
The finite time perspective of the third law is a statement on the
asymptotic rate of cooling as the absolute temperature is approached.
These restrictions are imposed on the optimal cooling rate. 
The behavior of the optimal cooling rate as the absolute 
temperature is approached is a third law upper bound on the cooling rate.
The main finding of this paper is that the optimal cooling rate converges to 
zero linearly with temperature, and the entropy production
reaches a constant when the cold bath temperature approaches absolute 
zero.

\section{Basic Assumptions and Formal Background for the 
Heat Engine and the Heat Pump}
\label{sec:engine}

Heat engines  and heat pumps are characterized by three attributes:  
the working medium, the cycle of operation, and the dynamics which govern 
the cycle. Heat baths by definition are large enough so that 
their temperatures is constant during the cycle of operation.
The heat engine and the heat pump are constructed from the same components
and differ only by their cycle of operation.  

\subsection{The Working Medium}
\label{subsec:medium}

The working medium consists of an ideal ensemble of 
many non-interacting discrete level systems.
Specifically, the analysis is carried out on two-level systems (TLS)
but an ensemble of harmonic oscillators  \cite{feldmann96}
would lead to equivalent results. 

The TLS systems are envisioned as spin-1/2 systems. The lack of spin-spin 
interactions enables the description of  the energy exchange 
between the working medium and the surroundings in terms of a single TLS. 
The state of the system is then defined by the average occupation 
probabilities $P_+$ and $P_-$ corresponding to the energies ${1 \over 2}
\omega$ and $-{ 1\over 2} \omega$, where $\omega$ is the energy gap
between the two levels.
The average  energy per spin is given  by
\begin{eqnarray}
E= {P_+} \cdot \left( {1  \over 2} \omega \right)
+ {P_-} \cdot \left( - {1  \over 2} \omega \right)
\label{energy}
\end{eqnarray}
The polarization, $S$, is defined by
\begin{eqnarray}
S~=~ {1  \over 2} (P_+ ~-~ P_-) ~~~,
\label{S}
\end{eqnarray} 
and thus the energy can be written as $E= \omega S$.
Energy change of the working medium can occur either by population transfer
from one level to the other (changing S) or by changing the energy gap between 
the two levels (changing $ \omega$). Hence
\begin{eqnarray}
dE ~ = ~ Sd \omega ~+~ \omega dS~~.
\label{dE}
\end{eqnarray}
Population transfer is the microscopic realization of heat exchange. 
The energy change due to external field variation is associated with work. 
Eq. (\ref{dE}) is therefore the first law of thermodynamics:
\begin{eqnarray}
D {\cal W} ~\equiv~ S d \omega ~~;~~ D {\cal Q} ~\equiv~ \omega dS~~.
\label{workheat}
\end{eqnarray}    
Finally, for TLS the internal temperature, $T^\prime$, is always defined
via the relation
\begin{eqnarray}
S~=~-{1  \over 2} \tanh \left( {{\omega} \over {2 k_B T^\prime}} \right)~~.
\label{Tinternal}
\end{eqnarray}
Note that the polarization $S$ is negative as long as the 
temperature is positive.

\subsection{The Cycle of Operation}
\label{subsec:cycle}
\subsubsection{Heat engines cycle}
\begin{figure}[tb]
\vspace{-0.66cm}
\hspace{3.cm}
\psfig{figure=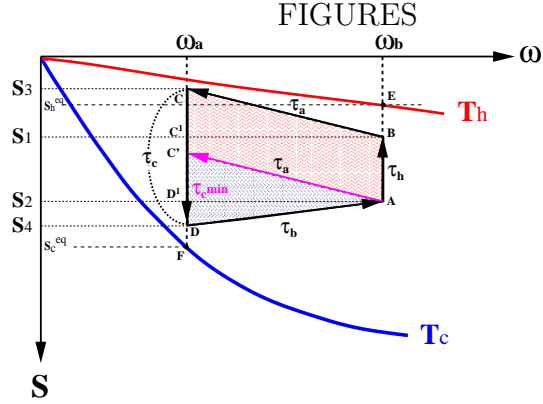,width=0.43\textwidth}
\vspace{0.5cm}
\caption{The heat engine with friction in the $\omega$~,$S$ plane.
$T_h$ is the hot  bath temperature.
$\tau_h$ is the time allocation when in contact with the hot bath.
$T_c$ and $\tau_c$ represent the temperature and time allocation
for the cold bath. $\tau_a$ represents the time allocation for compression
(field change from $\omega_b$ to $\omega_a$) and $\tau_b$ for expansion.    
The area~ A,B,~C$^1$,~D$^1$ is the positive work done by the system,
while the areas~ C,~C$^1$,~S$_1$,~S$_3$, and D$^1$,~D,~S$_4$,~S$_2$
represent the negative work done by the system.    }
\label{fig:cycle1}
\end{figure}

The cycle of operation is 
analyzed in terms of the polarization and frequency   $(S, \omega)$. 
A schematic display is shown  in Fig.(\ref{fig:cycle1}) for a constant total
cycle time, $ \tau$. 
The present engine is an irreversible four stroke 
engine \cite{feldmann96} resembling the Stirling cycle, 
with the addition of internal friction.  
The direction  of motion along  the cycle is  chosen such that net 
positive work is produced.

The four branches of the engine will be now briefly described.
 
On the first branch, 
$A \rightarrow B$, the working medium is coupled to the hot bath of temperature
$T_h$ for period $ \tau_h$, while the energy gap is kept fixed at the value
$ \omega_b$. The conditions are such that the internal temperature of the 
medium is lower than $T_h$.  In this branch, the polarization
is changing from the initial polarization $S_2$ to the polarization  $S_1$.
The inequality to be fulfilled is therefore:
\begin{eqnarray}
S_1< ~-{1  \over 2} \tanh \left( {{\omega_b} \over {2 k_B T_h}} \right)~~.
\label{ineqhb}
\end{eqnarray}
Since $ \omega$ is kept fixed, no work is done and the only energy transfer
is the heat $\omega_b (S_1-S_2)$ absorbed  by the working medium.
   
In the second branch,
$B \rightarrow C$ the working medium is decoupled from the hot bath
for a period $  \tau_a$, and the energy gap is varied linearly in time, 
from $ \omega_b$ to $ \omega_a$.  
In this branch work is done to overcome the inner friction 
which develops heat, 
causing the polarization to increase from $S_1$ to $S_3$ 
(Cf. Fig. \ref{fig:cycle1}).  
The change of the internal temperature is the result of two 
opposite contributes.
First lowering the energy gap leads to a lower inner temperature for constant 
polarization $S$. Second increase in polarization due to friction, leads to an
increase of the inner temperature for fixed $\omega$. 
The inner temperature  $T^\prime$ at point C might therefore be 
lower or higher 
than the initial temperature at point B. 
 
The third branch $C \rightarrow D$, is similar to the first. The 
working medium is now coupled to a cold bath at temperature $T_c$ for time 
$ \tau_c$. The polarization changes on this branch from $S_3$ to the 
polarization $S_4$. For the cycle to close, $S_4$ should be lower than $S_2$. 
At the end of the cycle the internal 
temperature of the working medium should be higher than the cold bath 
temperature, $T^\prime  > {T_c}$, leading to:

\begin{eqnarray}
S_4~ > ~-{1  \over 2} \tanh \left( {{\omega}_a \over {2 k_B T_c}} \right)~~.
\label{ineqcb}
\end{eqnarray}
Since $ S_4 <  S_1 $ (Fig. \ref{fig:cycle1}),  it follows from Eq.
( \ref{ineqhb}) and Eq.( \ref{ineqcb}), that:
\begin{eqnarray}
\left (\omega_a \over T_c \right)~ >~ \left (\omega_b \over T_h \right)
\label{ineqwt}
\end{eqnarray}
Inequality (\ref{ineqwt}) is equivalent to the Carnot efficiency bound, 
from  Eq. (\ref{ineqwt}) one gets:

\begin{eqnarray}
1~~-~~
\left (\omega_a \over \omega_b \right)~ <~1~~-~~
\left (T_c \over T_h \right)~=~ \eta_{Carnot}
\label{ineqwt1}
\end{eqnarray}
The present model is a quantum analogue of the
Stirling engine which also has Carnot's efficiency as an upper bound.

The polarization $S$ changes uni-directionally along the 'adiabats' 
due to the increase of the excited level
population as a result of the heat developed in the working fluid
when work is done against friction, irrespective of the direction 
of the field change.

The fourth branch  $D \rightarrow A$, closes the cycle and is similar to the
second. The working medium is decoupled from the cold bath. 
In a period $ \tau_b$ the
energy gap is changing back to its original value, $ \omega_b$. 
The polarization increases from $S_4$ to the original value $S_2$. 

\pagebreak

\begin{table}
\caption{Work and heat exchange along the branches of the heat engine 
with friction} 
\begin{center}
\begin{tabular}{||c|c|c||}
\tableline
branch & work+[work against friction] & heat \\
\tableline
$A \rightarrow  B$ & $0$ & $  \omega_b (S_1-S_2) $ \\  
\tableline
$B \rightarrow  C$ & $(\omega_a - \omega_b) (S_1 + \sigma^2/
(2 \tau_a) )$ ~+~
[ $ \sigma^2 (\omega_a + \omega_b)/(2 \tau_a) $ ] & $0$   \\
\tableline
$C \rightarrow  D$ &  $0$ & $  \omega_a ((S_2-S_1)- \sigma^2
(1/\tau_a +1  / \tau_b)) $ \\
\tableline
$D \rightarrow  A$ &  $(\omega_b - \omega_a) (S_2 - \sigma^2/(2 \tau_b))$ 
~+~ [ $ \sigma^2 (\omega_a + \omega_b)/(2 \tau_b) $ ] & $0$  \\
\tableline
\end{tabular}
\end{center}
\label{tab:cycle1}
\end{table}

\subsubsection{Refrigerator cycle}   

\begin{figure}[tb]
\vspace{0.2cm}
\psfig{figure=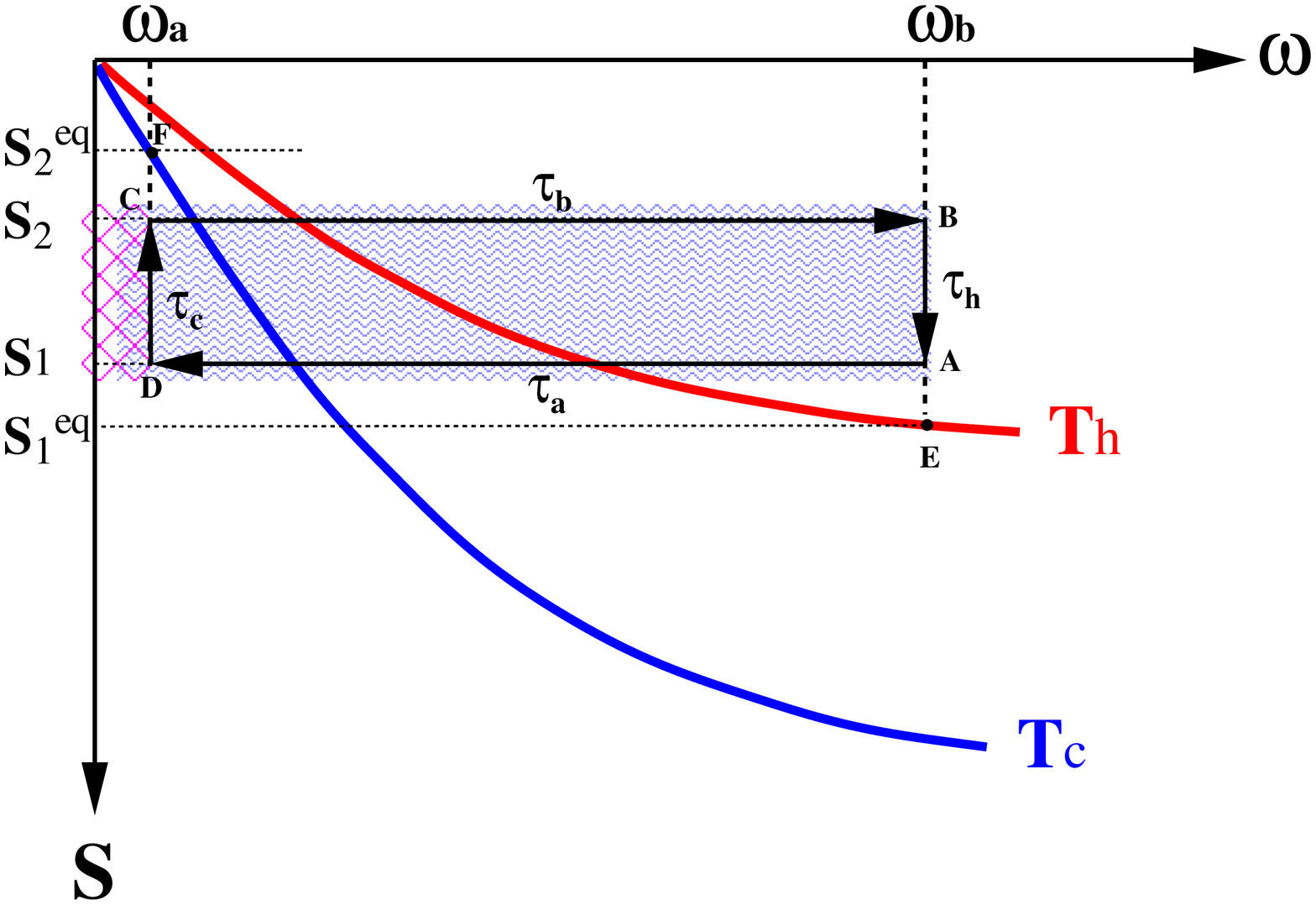,width=0.45\textwidth}
\vspace{-5.45cm}
\hspace{7.3cm}
\psfig{figure=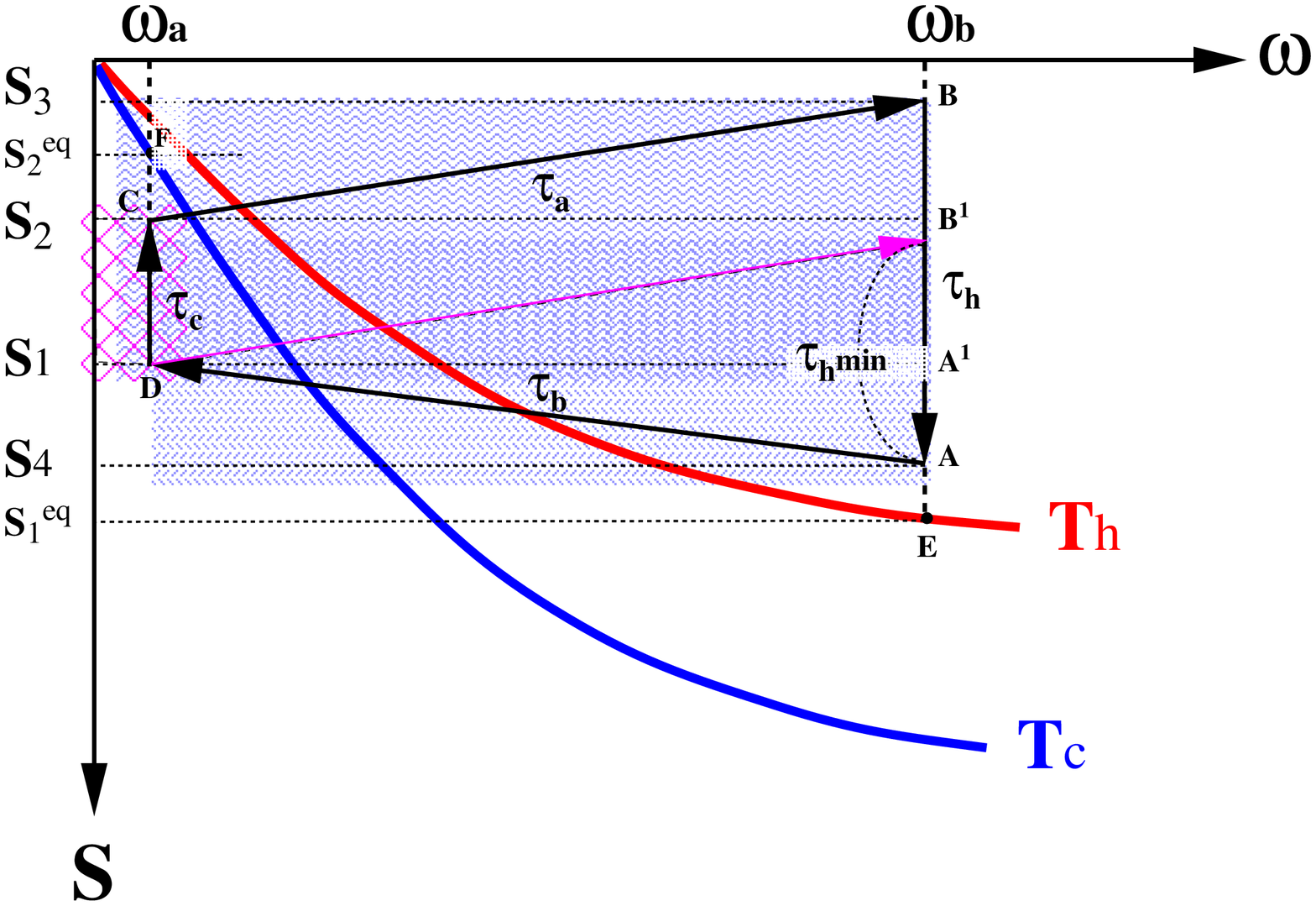,width=0.45\textwidth}
\vspace{0.4cm}
\caption{The cycle of operation of the heat pump. Left: without friction.
S$_1^{eq}$ is the hot bath equilibrium polarization. 
S$_2^{eq}$ is the cold bath equilibrium polarization.
The area enclosed by 
D,C,S$_2$ S$_1$ is the heat absorbed form the cold bath.
The area enclosed by DCBA is the work done on the system.
Right: with  friction. The area enclosed by 
D,C,S$_2$,S$_1$ is the heat absorbed form the cold bath. 
The work on the system is the area defined by the rectangles
B B$^1$ S$_2$ S$_3$ and  B$^1$ A$^1$ D C and A$^1$ A S$_4$ S$_1$.}
\label{fig:cycle2}
\end{figure}

The purpose of a heat pump is to  remove heat from the cold reservoir by  
employing external work.
The cycles of operation  in the  
$(S, \omega)$ plane is schematically shown in Fig. \ref{fig:cycle2},

The cycle of operation resembles the Ericsson refrigeration cycle \cite{chen98}.
The differences are in the dynamics of the microscopic working fluid
which are described in  subsection \ref{subsec:dynamics}. 
The work and heat transfer for the heat pump is
summarized in Table  \ref{tab:cycle}.

The four branches for the heat pump become:

In the first branch, 
$D \rightarrow C$, the working medium is coupled to the cold bath of 
temperature $T_c$ for time $ \tau_c$, while the energy gap is kept fixed 
at the value $ \omega_a$. The conditions are such that the internal 
temperature of the medium is lower than $T_c$ during $ \tau_c$. Along this 
branch, the polarization changes from the initial polarization $S_1$ 
to the polarization  $S_2$. 
Since $ \omega$ is kept fixed, no work is done and the only energy transfer
is the heat $\omega_a (S_2-S_1)$ absorbed  by the working medium. 
On this branch:
\begin{eqnarray}
S_2< ~-{1  \over 2} \tanh \left( {{\omega_a} \over {2 k_B T_c}} \right)~~.
\label{ineqcbr}
\end{eqnarray}

In the second branch,
$C \rightarrow B$ the working medium is decoupled from the cold bath, and
the energy gap is varied.  In the frictionless case the polarization $S_2$
is constant (Left of Fig. \ref{fig:cycle2}). 
The only energy exchange is the work done on  the system  
( Table  \ref{tab:cycle}).
When friction is added 
the polarization is changing from $S_2$ to $S_3$ in a period $  \tau_a$.
The energy gap  changes from $ \omega_a$ to 
$ \omega_b$ (Right of Fig. \ref{fig:cycle2}), according to a linear law. 
In addition to work,  heat 
is developing as a result of  the inner friction ( Table  
\ref{tab:cycle}).

The third branch $B \rightarrow A$, is similar to the first. The 
working medium is coupled to the hot bath at temperature $T_h$, for time 
$ \tau_h$, keeping the energy gap $ \omega_b$ fixed. In this branch the 
polarization changes from $S_2$ to  $S_1$ in the frictionless case, 
and from  $S_3$ to $S_4$ when friction is added.  
The constraint is that the 
internal temperature of the working medium should be higher than the 
hot bath temperature during the time  $ \tau_h$, $T^\prime > T_h$,
leading to the inequality (Fig. \ref{fig:cycle2}),
\begin{eqnarray}
S_1>~S_4~>~ ~-{1  \over 2} \tanh \left( {{\omega_b} \over {2 k_B T_h}} 
\right)~~.
\label{ineqhbr}
\end{eqnarray}
therefore  $ S_2 > S_1 $. From Eqs.
(\ref{ineqcbr}) and (\ref{ineqhbr}), the condition for the interrelation
between the bath temperatures and the field values becomes:
\begin{eqnarray}
\left (\omega_a \over T_c \right)~ <~ \left (\omega_b \over T_h \right)
\label{ineqwtr}
\end{eqnarray} 
which is just the opposite inequality of the heat engine,
(Eq. \ref{ineqwt}).  In the heat pump
work is done {\em on} the  working fluid and since 
no useful work is done Carnot's bound is not violated.
  
The fourth branch  $A \rightarrow D$, closes the cycle and is similar to the
second. The working medium is decoupled from the cold bath, and the
energy gap changes back, during a period $ \tau_b$ 
to its original value, $ \omega_b$.

The results are summarized in  Table  \ref{tab:cycle}.

\begin{table}
\caption{Work and heat exchange along the branches of the 
heat pump without/with friction.} 
\vspace{0.3cm}
{\centering \begin{tabular}{|c|c|c|}
branch&
frictionless work/work{+}[work against friction]&
heat\\
\hline 
\hline 
D\( \rightarrow  \)C&
0&
\( \omega _{a}(S_{2}-S_{1}) \)\\
\hline 
C\( \rightarrow  \)B&
\multicolumn{1}{c|}{\( \begin{array}{c}
(\omega_{b}-\omega_{a})S_{2}\\
\left(\omega_{b}-\omega_{a})(S_{2}+\sigma^{2}/(2\tau_{a})\right)
+[\sigma^{2}(\omega _{a}+\omega_{b})/(2\tau_{a})]\\
\\
\end{array} \)}&
0\\
\hline 
B\( \rightarrow  \)A&
0&
\( \begin{array}{c}
\omega _{b}(S_{1}-S_{2})\\
\omega _{b}\left((S_{1}-S_{2})-\sigma ^{2}(1/\tau_{a}+1/\tau_{b})\right)\\

\end{array} \)\\
\hline 
A\( \rightarrow  \)D&
\( \begin{array}{c}
(\omega_{a}-\omega_{b})S_{1}\\
(\omega_{a}-\omega_{b})\left(S_{1}-\sigma^{2}/(2\tau_{b})\right)
+[\sigma ^{2}(\omega_{a}+\omega _{b})/(2\tau_{b})]\\
\\
\end{array} \)&
0\\
\end{tabular}\par}
\vspace{0.3cm}
\label{tab:cycle}
\end{table}

\subsection{Dynamics of the working medium}
\label{subsec:dynamics}

The dynamics of the system along the heat exchange branches 
is represented by changes in the level population of the two-level-system.
This is a reduced description in which the dynamical response of 
the bath is cast in kinetic terms \cite{geva2}.
Since the dynamics has been  described previously
\cite{feldmann96} only a brief summary of the main points 
is presented here, emphasizing the differences in 
the energy exchanges on the 'adiabats'.

\subsubsection{The dynamics of the heat exchange branches}  

The dynamics of the population at the two levels,  $P_+$ and $P_-$, are 
described via a master equation
\begin{eqnarray}
\left\{
\begin{array}{l}
{{{d P_+} \over {dt}} = -k_\downarrow P_+ ~+~ k_\uparrow P_-}\\
{{{d P_-} \over {dt}} = ~k_\downarrow P_+ ~-~ k_\uparrow P_-} 
\end{array}
\right.~~,
\label{master}
\end{eqnarray}
where $k_\downarrow$ and  $k_\uparrow$
are the transition rates from the  upper to the lower level and
vice versa. The explicit form of these coefficients depend on the nature
of the bath and the system bath coupling interactions.
The thermodynamics partition between system and bath is consistent with 
a weak coupling assumption \cite{geva2}.
Temperature enters through detailed balance. 
The equation of motion for the polarization $S$ obtained from
Eq. (\ref{master}) becomes: 
\begin{eqnarray}
{{dS} \over {dt}}~=~ - \Gamma (S ~-~ S^{eq})~~
\label{dSdt} 
\end{eqnarray}
where 
\begin{eqnarray}
\Gamma~=~  {k_\downarrow}~+~ {k_\uparrow}
\end{eqnarray}
and
\begin{eqnarray}
S^{eq}~=~-{1 \over 2} {{{k_\downarrow}~-~ {k_\uparrow}} \over
                       {{k_\downarrow}~+~ {k_\uparrow}}}
= -{1 \over 2} \tanh \left( {\omega \over {2 k_B T}} \right)~~   
\label{Seq}
\end{eqnarray}
where $S^{eq}$ is the corresponding equilibrium polarization.
It should be noticed that in a TLS there is a one to one correspondence between
temperature and polarization thus internal temperature is well defined even 
for non-equilibrium situations.

The general solution of Eq (\ref{dSdt}) is,
\begin{eqnarray}
S(t) ~=~ S^{eq} ~+~ ( S(0)~-~ S^{eq}) e^{- \Gamma t}~~.
\label{Sgeneral}
\end{eqnarray} 
where S(0) is the polarization at the beginning of the branch.

From Eqs. (\ref{dSdt}) and  (\ref{Seq}) the rate of heat change
becomes: 
\begin{eqnarray}
\dot{\cal Q}~~=~~  \omega \dot{S}
\label{Qdot}
\end{eqnarray}  
See also \cite{geva0}.

For convenience, new  time variables are defined:
\begin{eqnarray} 
x~=~e^{-\Gamma_c \tau_c }~~ , ~~y~=~e^{-\Gamma_h \tau_h}
\label{xy}
\end{eqnarray}
These expressions represent a nonlinear mapping of the time allocated
to the hot and cold branches by the heat conductivity $\Gamma$. 
As a result, the time allocation and the heat conductivity parameter
become dependent on each other.

Figure \ref{fig:cycle1} and  \ref{fig:cycle2} show
that the friction induces an asymmetry between the time allocated to
the hot and cold branches since more heat has to be dissipated 
on the cold branch.

\subsubsection{The dynamics on the 'adiabats'}  

The external field $\omega$ and its rate of change $\dot \omega$ 
are control parameters of the engine.
For simplicity it is assumed that the  field  changes linearly with time:
\begin{eqnarray}
{\omega}(t)~~=~~\dot{\omega} t~~+~~\omega(0)
\label{omegat}
\end{eqnarray}

Rapid change in the field causes non-adiabatic behavior which to lowest
order is proportional to the rate of change $\dot \omega$.
In this context non-adiabatic is understood in its quantum mechanical
meaning.
Any realistic assumption beyond the ideal non-interacting TLS will lead to
such non-adiabatic behavior.
It is therefore assumed that the phenomena can be described
by a  friction coefficient $\sigma$ which forces 
a constant speed  polarization change $\dot{S}$: 
\begin{eqnarray}
\dot{S}~~=~~  \left( { \sigma} \over {t^ \prime} \right)^ 2
\label{Sdotadiab}
\end{eqnarray}
where $t^ \prime$ is the time allocated to the corresponding 'adiabat'.
Therefore, the polarization as a function of time becomes:
\begin{eqnarray}
S(t)~~=~~ S(0)~~+~~  \left( { \sigma} \over {t^ \prime} \right)^ 2 t
\label{Sadiab}
\end{eqnarray}
where     $t \geq 0$ ,  $t  \leq t^ \prime$. 
A modeling assumption of internally dissipative friction, similar 
to Eq.(\ref{Sdotadiab}), was also made by 
Gordon and Huleihil (\cite{gordon91}).
Friction does not operate on the heat-exchange branches,  
there is no nonadiabtic effect since the fields
$\omega_a$ and $\omega_b$ are constant in time.
The irreversibilities on those branches are due to the
transition rates ($\Gamma$) of the master equation.

From Fig.(\ref{fig:cycle1}),  Eq. (\ref{workheat}), and Eq. (\ref{Sadiab}) 
the polarization, for the $B \rightarrow C$ branch of the heat engine becomes: 
\begin{eqnarray}
S_C=S_3~~=~~ S_1~~+~~ \left( {  \sigma^2 }  \over { \tau_a }~
\right)~.
\label{SadBC}
\end{eqnarray}
The work done on this branch is:
\begin{eqnarray}
{\cal W}_{BC} ~=~ \int_{0}^{\tau_a} D {\cal W} ~=~
\int_{0}^{\tau_a}{S \dot{\omega}dt}~~
=~{(\omega_a~-~\omega_b)}\left( S_1 ~+~{1 
\over 2} \left( {\sigma}^2 \over 
{\tau_a} \right) \right)
\label{WadBC}
\end{eqnarray}
The heat generated on this branch in the working fluid, which 
is the work against the friction, becomes:
\begin{eqnarray}
{\cal Q}_{BC} ~=~ \int_{0}^{\tau_a} D {\cal Q} ~=~
\int_{0}^{\tau_a}{\omega \dot{S}dt}~~
=~{ \sigma^2  (\omega_a+\omega_b)  
\over { 2 \tau_a} }   
\label{QadBC}
\end{eqnarray}
This work is dependent on the friction coefficient and inversely 
on the time allocated to the 'adiabats'.
The computation  for the other branches of  the heat engine and
heat pump are similar.

\subsubsection{Explicit expressions for the  polarizations imposed by the closing of 
the cycle.}  

By forcing the cycle to close, the four corners of the cycle
observed in Fig. \ref{fig:cycle1} are linked.
Applying Eq.  (\ref{Sgeneral})  leads to the equations:
\begin{eqnarray}
\begin{array}{l}
S_1~=~S_2 y~+~ S_h^{eq} (1-y)\\
S_3~=~S_1~+~{  \sigma^2 \over \tau_a } \\
S_4~=~S_3 x~+~ S_c^{eq}(1-x)\\
S_2~=~S_4~+~{  \sigma^2 \over \tau_b }
\end{array} 
\label{S2xy}
\end{eqnarray}
The solutions for $S_1$, $S_2$ and $S_1-S_2$ are
\begin{eqnarray}
\begin{array}{l}
S_1= S_c^{eq}~+~{ \Delta S^{eq} (1-y)~+~ \sigma^2 y G(x) \over (1-xy)  }~~
=~~  S_h^{eq}~-~{ \Delta S^{eq} y(1-x)~-~ \sigma^2 y G(x) \over (1-xy)  }\\
S_2= S_c^{eq}~+~{ \Delta S^{eq} x(1-y)~+~ \sigma^2 G(x) \over (1-xy)  }~~
=~~  S_h^{eq}~-~{ \Delta S^{eq} (1-x)~-~ \sigma^2  G(x) \over (1-xy)  }
\end{array}
\label{eqS2}
\end{eqnarray}
and
\begin{eqnarray}
S_1-S_2=  ( \Delta S^{eq}) F(x,y) ~-~{   \sigma^2 (1-y) G(x) 
\over (1-xy)  } 
\label{eqS1S2}
\end{eqnarray}
where 
\begin{eqnarray}
\nonumber
F(x,y)= {  (1-x) (1-y) \over (1-xy)  } ~~~,~~~
 \Delta S^{eq}~~=~~(S_h^{eq}- S_c^{eq}) 
~~~,~~~
G(x)~=~ (x/\tau_a + 1/\tau_b)
\end{eqnarray}
The constraint that the cycle must close leads to conditions on the
polarizations $S_1$ and $S_2$ and on the minimum cycle time $\tau_{c,min}$.
Eqs. (\ref{eqS2}) shows that both $S_1$ and $S_2$
are bounded by $S_h^{eq}$ and $S_c^{eq}$. 
The minimum cycle time is obtained when the polarizations coincide
with the hot bath polarization: $S_1$=$S_2$=$S_h^{eq}$.
In this  case, $\tau_h$=0, and 
from Eqs. (\ref{xy}) and  (\ref{eqS1S2}) 
the minimum time allocation on the cold bath $\tau_{c,min}$ is computed,
\begin{eqnarray}
x_{max} ~=~{  (S_h^{eq}- S_c^{eq})  ~-~  
        { \sigma^2 / \tau_b }  
\over 
  (S_h^{eq}- S_c^{eq})  ~+~  
        { \sigma^2 / \tau_a } }
\label{minxc}
\end{eqnarray}
or
\begin{eqnarray}
\tau_{c,min} ~=~ -{1/ \Gamma_c} \lg { (S_h^{eq}- S_c^{eq})  ~-~  
        { \sigma^2 / \tau_b } 
            \over 
 { (S_h^{eq}- S_c^{eq})  ~+~  
        { \sigma^2 / \tau_a }}  }
\label{mintauc}
\end{eqnarray}
From this  expression for $\tau_{c,min}$ the lower bound for the overall
cycle time, is obtained (The left of Fig.{ \ref{fig:taumin7}}) : 
\begin{eqnarray}
\tau ~ \geq ~ \tau_{min} ~=~ \tau_{c,min} + \tau_a + \tau_b 
\label{mintau1}
\end{eqnarray}
When the  minimum cycle time Eq. (\ref{mintauc}) diverges,
the cycle cannot be closed. This condition imposes
an upper bound on the friction coefficient  $\sigma$
\begin{eqnarray}
\sigma ~~\leq~~\sigma^{up} ~=~ \sqrt{~\tau_b  (S_h^{eq}- S_c^{eq})}.  
\label{sigup}
\end{eqnarray}
or
\begin{eqnarray}
~\tau_b > \tau_{b,min}~=~{ \sigma^2 \over (S_h^{eq}- S_c^{eq}) }.
\label{tbmin}
\end{eqnarray}

Closing of the cycle imposes similar constraints on the 
minimal cycle time under friction for the heat pump. 
The value of the polarization difference $S_2-S_1$ using the 
notation of Fig. \ref{fig:cycle2} becomes:
\begin{eqnarray}
S_2-S_1=  ( S_2^{eq}- S_1^{eq}) F(x,y) ~-~  
        {  \sigma^2  (1-x)  (y/\tau_a + 1/\tau_b) 
            \over (1-xy)  } 
\label{eqS2S1}
\end{eqnarray}
The minimum cycle time is calculated in
the limit when $\tau_c$=0, leading to $S_2$=$S_1$=$S_2^{eq}$.
From Eqs. (\ref{xy}) and  (\ref{eqS2S1}) the minimum time allocation on the
hot branch $\tau_{h,min}$ is computed: 
\begin{eqnarray}
y_{max} ~=~{  (S_2^{eq}- S_1^{eq})  ~-~  
        { \sigma^2 / \tau_b }  
\over 
  (S_2^{eq}- S_1^{eq})  ~+~  
        { \sigma^2 / \tau_a } }
\label{maxyh}
\end{eqnarray}

\begin{eqnarray}
\tau_{h,min} ~=~ -{1/ \Gamma_h} \lg { (S_2^{eq}- S_1^{eq})  ~-~  
        { \sigma^2 / \tau_b } 
            \over 
 { (S_2^{eq}- S_1^{eq})  ~+~  
        { \sigma^2 / \tau_a }}  },
\label{mintauh}
\end{eqnarray}
where  $S_2^{eq} $ is point  F  and  $S_1^{eq}$ is point E on Fig. 
\ref{fig:cycle2}.
Using $\tau_{h,min}$ the lower bound for the overall
cycle time, is computed
\begin{eqnarray}
\tau ~ \geq ~ \tau_{min} ~=~ \tau_{h,min} + \tau_a + \tau_b 
\label{mintau2}
\end{eqnarray}
Closing the cycle imposes a minimum cycle time
for both the heat engine and the heat pump, 
which is a monotonically increasing function of the friction 
coefficient $\sigma$. The divergence of $\tau_{min}$ 
imposes a maximum value for the friction coefficient $\sigma$.

\subsection{Finite Time Analysis}
\label{sec:FTT}

\subsubsection{Quantities to be Optimized.}
\label{subsec:optquant}

The primary variable to be optimized is the power of the heat engine 
and the heat-flow extracted from the cold reservoir of the heat pump. 
For a preset cycle time, optimization of the power is equivalent to
optimization of the total work, while optimization of heat flow is
equivalent to the optimization of the heat absorbed.
The entropy production will also be analyzed.

(1) \bf The total work done on the environment per cycle of the 
Heat Engine. \rm

The total work of the engine, is the sum of the work on each branch:
Cf. (Table \ref{tab:cycle1} and  Fig. \ref{fig:cycle1}):
\begin{eqnarray}
{\cal W}_{cyle1} ~=~ \oint D {\cal W} ~=~
-\left( W_{AB}+W_{BC}+W_{CD}+W_{DA} \right) 
\label{WcycH1}
\end{eqnarray}
which becomes: 
\begin{eqnarray}
{\cal W}_{cyle1} ~=~ (\omega_b-\omega_a)
 (S_1 - S_2)~-~ \sigma^2 \omega_a ( 1/ \tau_a ~+~ 1/ \tau_b ) 
\label{WcycH11}
\end{eqnarray}
The negative sign is due to the  convention of positive ${\cal W}$ 
when work is done on the system.  

Analyzing Eq. (\ref{WcycH11}), the work  is partitioned into
three positive and negative areas. The positive area (left rotation)
\begin{eqnarray}
{\cal W}_p~=~  (\omega_b-\omega_a) (S_1 - S_2) 
\label{area1}
\end{eqnarray}
is defined by the points $ A,B,C^1,D^1 $ in  Fig.{ \ref{fig:cycle1}}.
The two negative areas (right rotation)
\begin{eqnarray}
{\cal W}_n~=~ \sigma^2 \omega_a ( 1/ \tau_a )~~+~~ \sigma^2 \omega_a ( 1/ \tau_b )
\label{area23}
\end{eqnarray}
are defined by the points $ C,C^1,S_1,S_3$ and $D^1,D,S_4,S_2$
in  Fig.{ \ref{fig:cycle1}}.

The cycle which achieves the minimum cycle time
$ \tau$~=~$ \tau_{c,min}$, produces zero positive work ${\cal W}_p=0$.
The corners A and B coincide at E, and $C^1$ coincides with
$D^1$. The negative work of   Eq. (\ref{area23}), is defined by the
corners $C,D,S_4,S_3$ and is 'cut' by the $S_h^{eq}$ line
(Cf. the right of Fig. \ref{fig:powmin0}).
The cycle has  negative total work,
meaning that work is done {\em on } the working fluid 
against friction. 
When $ \tau$ increases
beyond ~$ \tau_{c,min}$ , $S_1$ diverts 
from  $S_2$, becoming lower than $S_h^{eq}$~~ 
(Cf.Eq. (\ref{eqS2})).
At a certain point, the work done against friction
is exactly balanced by the useful work of the engine.
The minimum time in which this balance is achieved is designated $ \tau_0$.
Its value which can be deduced from Eq.  (\ref{WcycH11}) 
is worked out in appendix \ref{sec:Apptau0}.

\begin{figure}[tb]
\vspace{-.5cm}
\psfig{figure=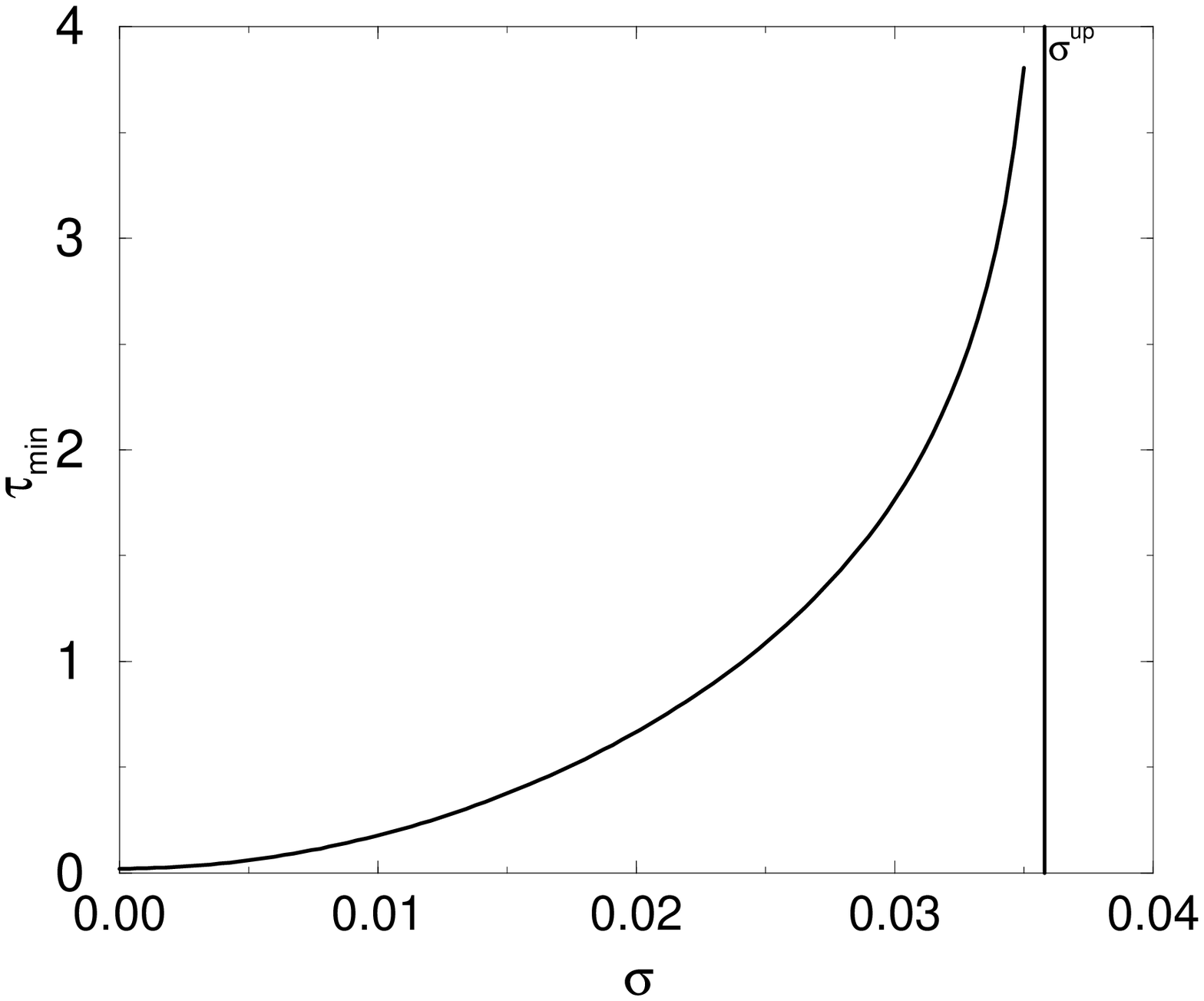,width=0.5\textwidth}
\vspace{-6.78cm}
\hspace{7.cm}
\psfig{figure=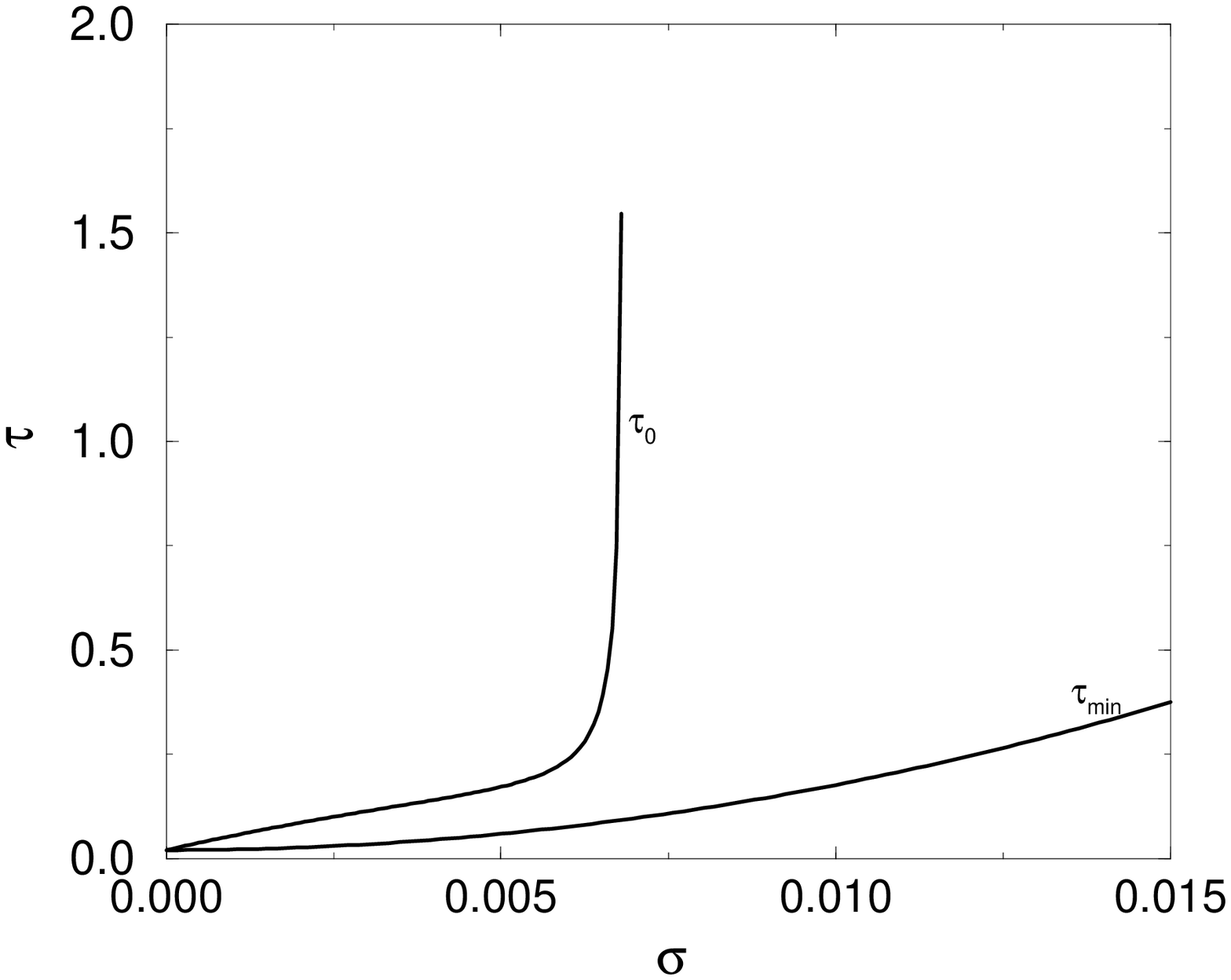,width=0.5\textwidth}
\vspace{0.1cm}
\caption{Left: Minimal cycle time $\tau_{min}$ as a function 
of the inner friction parameter $\sigma$ for the heat engine.
The vertical line represents  the upper-bound of $\sigma$.
Dimensionless units are used in which $k_b=1$ and $\hbar=1$.
The parameters used are: $\omega_a=1794$, $\omega_b=4238$,
$T_c=500$, $T_h=2500$, $\Gamma_c=1$ and $\Gamma_h=2$.
Right: Comparison between $\tau_{min}$ and $ \tau_0$,
the minimum cycle time for power production.   }
\label{fig:taumin7}
\end{figure}

\begin{figure}[tb]
\vspace{-.7cm}
\psfig{figure=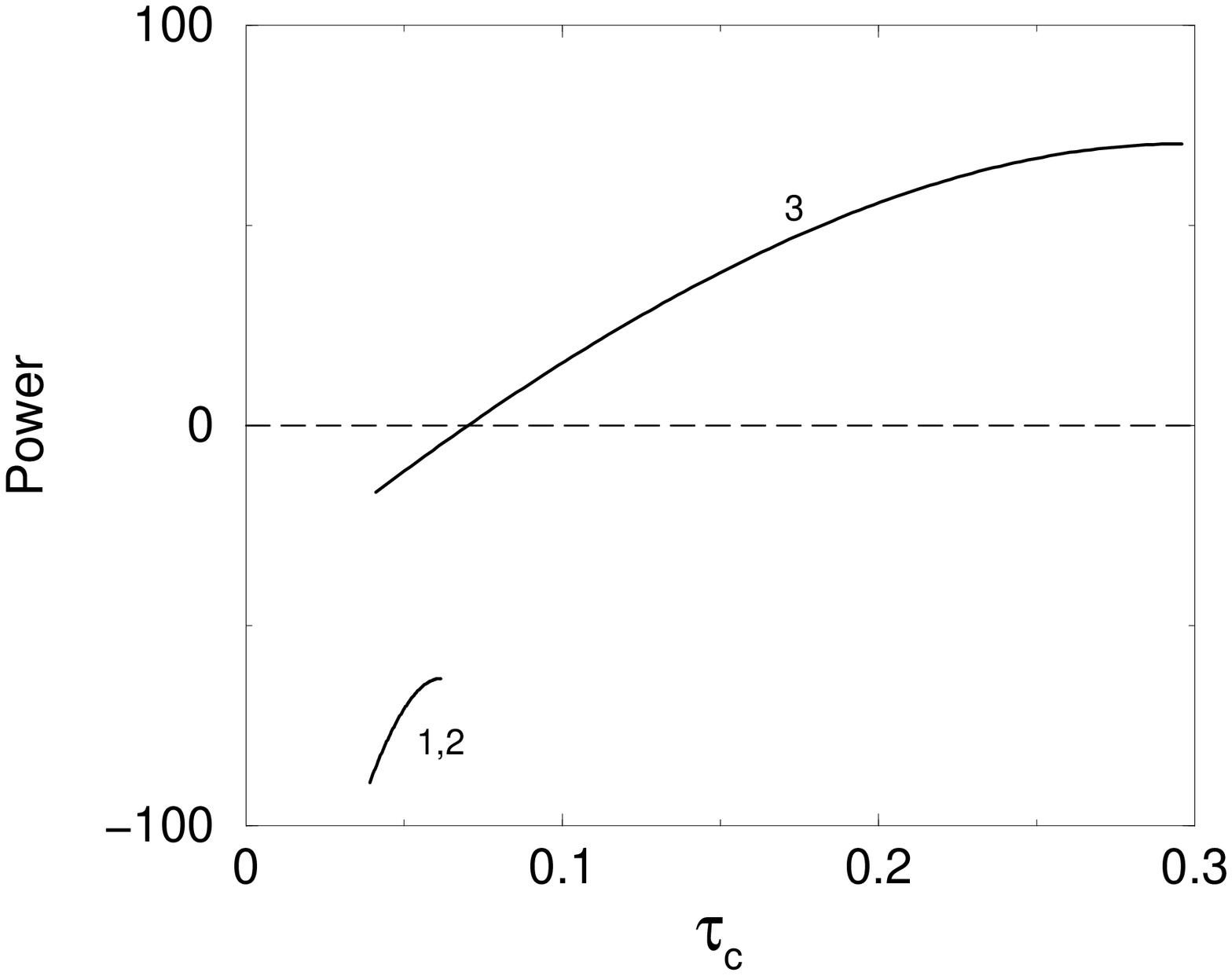,width=0.5\textwidth}
\vspace{-6.74cm}
\hspace{7.cm}
\psfig{figure=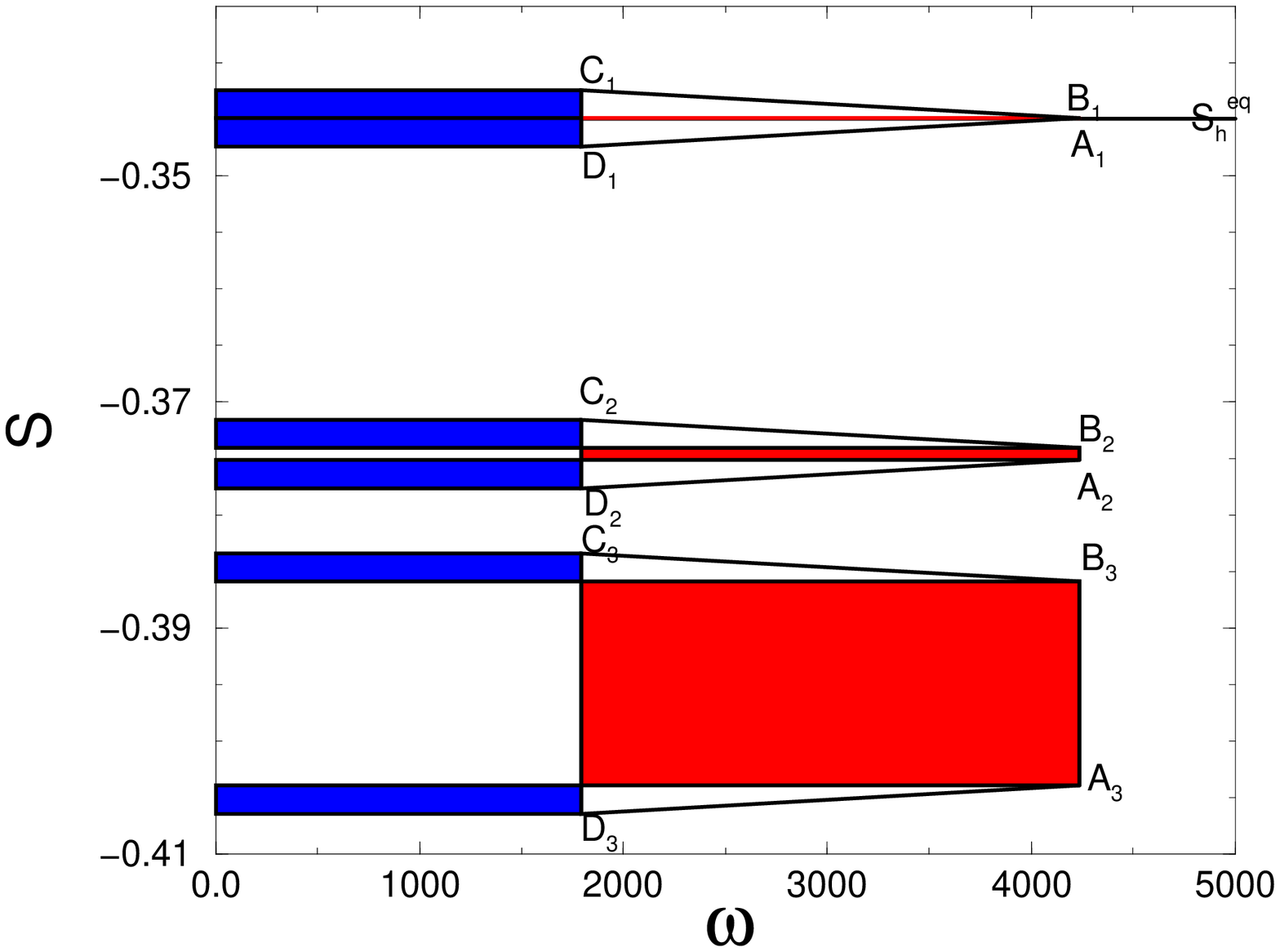,width=0.5\textwidth}
\vspace{0.1cm}
\caption{Left: Power as a function of the time allocation
on the cold branch corresponding to the friction coefficient
$ \sigma= 0.005$  with changing cycle times.
The cycle time values are:
for curve 1, $ \tau$ = $ \tau_{min} $ =0.059, for curve 2,
 $ \tau=0.1$ (the first two plots overlap) and for
curve 3,  $ \tau=0.5$. Other parameters are the same as in figure 3.
The dashed horizontal line is the line of zero power.  
Right:  The cycles corresponding to the power plots. 
Negative work is in blue and positive work is in red.
Note that for cycles 1 and 2,
the total area is negative and, therefore, the power output is negative.}
\label{fig:powmin0}
\end{figure}

The minimum cycle time   $ \tau_{min}$ is compared to $ \tau_0$, 
the minimum time needed to obtain positive power shown in the right of
Fig.{ \ref{fig:taumin7}} as a function of the friction $ \sigma$.
Both functions increase with friction,
but $\tau_0$ diverges at a much lower friction parameter.
Above this friction parameter no useful work can be obtained from the engine.
The divergence of $ \tau_{min}$ corresponds to a larger friction value
where the cycle cannot be closed.

\begin{figure}[tb]
\vspace{1.2cm}
\psfig{figure=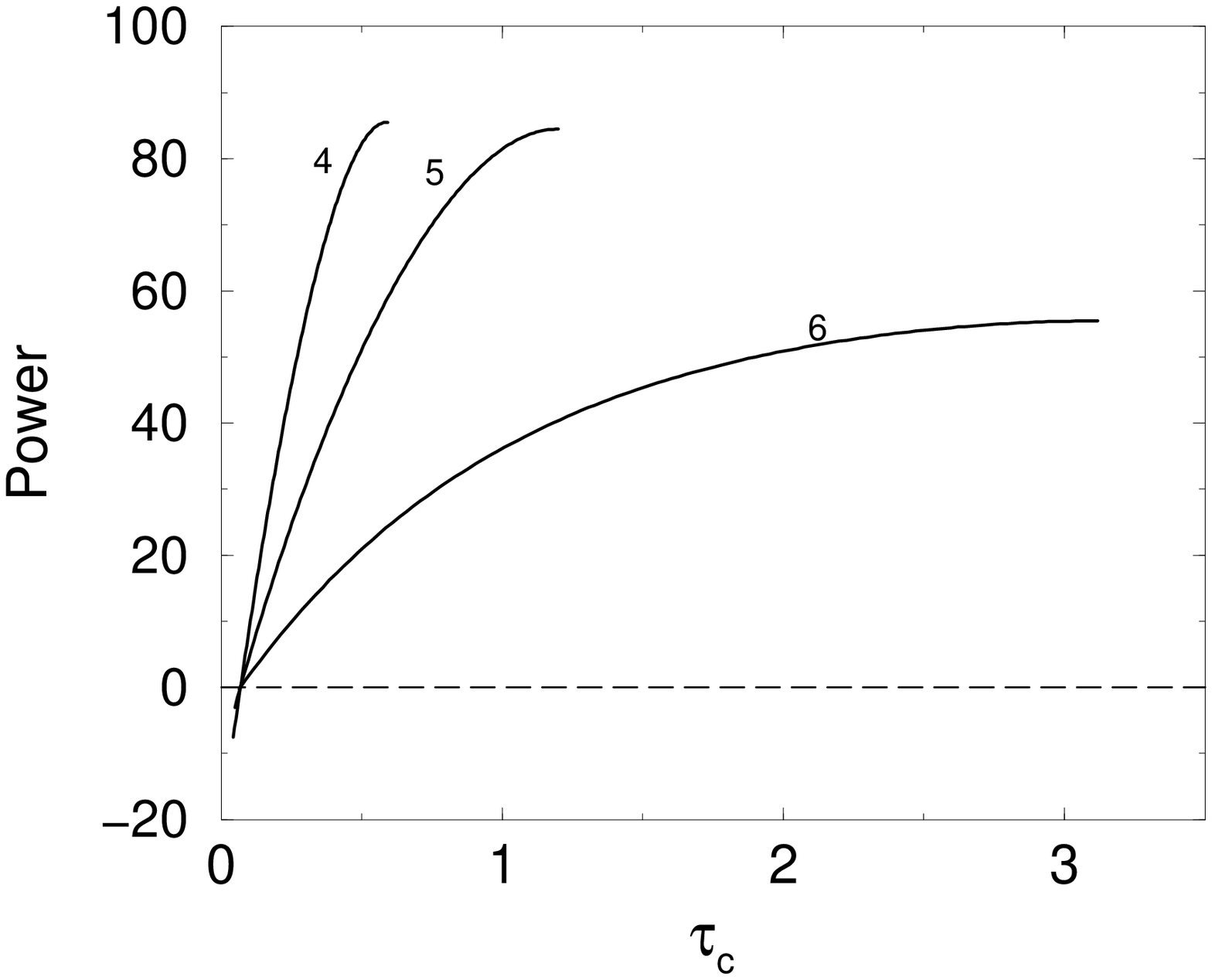,width=0.5\textwidth}
\vspace{-6.78cm}
\hspace{7.cm}
\psfig{figure=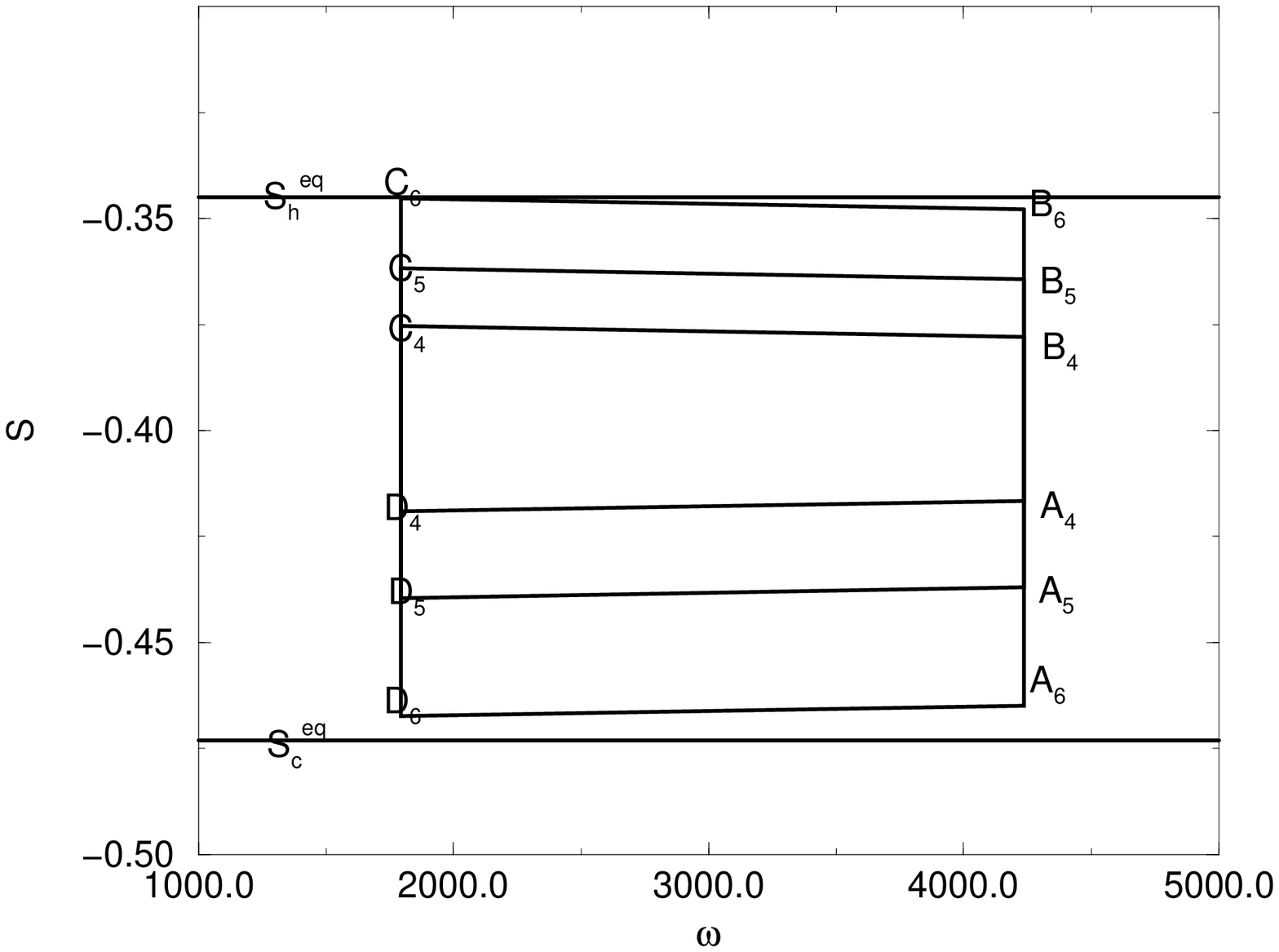,width=0.5\textwidth}
\vspace{0.1cm}
\caption{
Left: Power as a function of the time allocation
on the cold branch corresponding to the friction coefficient
$ \sigma= 0.005$  with changing cycle times.
The cycle time values are:
for curve 4, $ \tau=1$, for curve 5,  $ \tau=2$ and for curve 6, 
$ \tau=5 $. 
The dashed horizontal line is the line of zero power.
Right:  The cycles corresponding to the power plots.
All the constant parameters are as in  Fig. \ref{fig:powmin0} }
\label{fig:powerc3}
\end{figure}

When the total time allocation is sufficient, i.e.
$ \tau~>~\tau_0$, work is done on the environment, and
$S_1$ starts to increase. For long cycle times $S_1$ will approach 
$S_h^{eq}$, while $S_2$ will approach  $S_c^{eq}$.
The constant negative area will become negligible in comparison
to the positive area ( Fig.{ \ref{fig:powerc3}}).  

To study the influence of friction on the work output the polarization 
difference from Eq. (\ref{eqS1S2})    S$_1$-S$_2$ is inserted into
the work expression Eq.  (\ref{WcycH11}), leading to:
\begin{eqnarray}
{\cal W}_{cyle1} ~=~(\omega_b-\omega_a) 
          (S_h^{eq}-S_c^{eq})F(x,y)~-~{\cal W}_{\sigma1}
\label{WcycH2}
\end{eqnarray}
where
\begin{eqnarray}
{\cal W}_{\sigma1}~=~
\sigma^2 \left(
{ {\omega_b(1-y)} (x/\tau_a~+~1/\tau_b) \over {1-xy} } ~+~
{ {\omega_a(1-x)} (1/\tau_a~+~y/\tau_b) \over {1-xy} }  \right) 
\label{WcycH3}
\end{eqnarray}
${\cal W}_{\sigma1}$ is the additional 'cost' due to friction
and  is always positive. 

The emergence of positive power $\cal P$ is shown in Fig. \ref{fig:powmin0}.
For a fixed cycle time the optimization of work is equivalent to
the optimization of power.

The first two cycles have a cycle time shorter than $\tau_0$,
and therefore do not produce useful work.
For cycle 3,  $\tau > \tau_0$ and positive work is obtained when
the time allocation on the cold bath is sufficient $ \tau_c~\geq~0.08$.

For longer total cycle times, the ratio between the negative area 
to the positive area decreases as can be seen in
Fig. \ref{fig:powerc3}. 

The position of the cycles in the $S$, $\omega$ coordinates relative to
$S_h^{eq}$ and $S_c^{eq}$ changes as a function of the cycle time.
Insight to the origin of the behavior of the 'moving' cycles is 
presented in Fig. \ref{fig:s1s2t1} of Appendix \ref{sec:MovingCycles}.

The calculation of the total work done on the 
working fluid per cycle, 
 ${\cal W}^{on}_{cycle3}$ for the
heat pump is described in appendix \ref{totalw}. 
See also ( Cf. Table (\ref{tab:cycle}) and  Cf. Fig. \ref{fig:cycle2}).

(2)  \bf The heat-flow(${\cal Q}_F$) \rm

The heat-flow, ${\cal Q}_F$, extracted from the cold 
reservoir is:
\begin{eqnarray}
~~~{\cal Q}_F~~~= \omega_a (S_2-S_1)/ \tau 
\label{rheatf0}
\end{eqnarray}

Due to the dependence of ${\cal Q}_F$ only on $S_2$-$S_1$, the cycle
is similar to the  cycle of the heat engine.

(3) \bf The entropy production~($\Delta {\cal S}^u$). \rm

The entropy production of the universe,~$\Delta {\cal S}^u$,  is
concentrated on the boundaries with the  baths since,
for a closed cycle, the  entropy of the working fluid  is constant.
The computational details for both the heat engine and the heat pump
are shown in appendix \ref{sec:entrfricA}. 
The entropy production and the power have a reciprocal relation
(See  Fig. \ref{fig:entr1}). For example, the entropy production 
increases with $\sigma$, while the power decreases.

(4)  \bf Efficiency. \rm

The efficiency of the heat engine is the ratio of useful work to 
the heat extracted from the hot bath.
\begin{equation} 
{\eta}_{H.E.}~=~~{ {{\cal W}_{cycle}} \over {{\cal Q}_{absorbed}} }~~=~~
  \eta^{fricles}_{H.E.}~~-~~
\left( {{ \sigma^2 \omega_a (1/\tau_a+1/\tau_b)} \over 
\omega_b ( S_1~-~S_2 )  }  \right) 
\label{etahe}
\end{equation}

where $  \eta^{fricles}_{H.E.}~=~ ~{(1~-~{ \omega_a / \omega_b})}$

When the cycle time approaches its minimum
${\tau}~~\rightarrow~~{ \tau_{min}},$ 
the efficiency diverges: $ ~~{\eta}_{H.E.}~~{ \longrightarrow}~~-~{ \infty}$.
The efficiency becomes positive only when ${\tau}~~\geq~~{ \tau_{0}}$.
Using Eq. (\ref{etahe}) a bound for the efficiency is obtained:
\begin{equation} 
   0 ~~ < ~~ {\eta}_{H.E.}~~\leq~~{ \eta^{fricles}_{H.E.}
~~-{T_c \over {T_h}}}
\left( {{  
~~   \sigma^2  (1/\tau_a+1/\tau_b)} \over 
 ( S_1~-~S_2 )  }  \right) 
\label{etahe3}
\end{equation}

The cooling efficiency of the refrigerator will be:
\begin{equation} 
{\eta}_{Rf}~=~~{ {\cal Q}_{DC}  \over  {\cal W}_{cycle}^{on}  }~=~
{ \omega_a (S_2-S_1)  \over \left( (\omega_b-\omega_a) (S_2-S_1)
~+~  \sigma^2 \omega_b (1/\tau_a+1/\tau_b) \right) } 
\label{etafr4}
\end{equation}                                                                
or:
\begin{equation} 
{ 1 \over \eta_{Rf} } + 1~=~{ 1 \over COP } + 1~=
~{ \omega_b \over \omega_a } \left( 1+{ \sigma^2 (1/ \tau_a
+1/ \tau_b) \over {  (S_2-S_1) }} \right) 
~>~{ T_h \over T_c } \left( 1 + { \sigma^2 (1/ \tau_a
 + 1/ \tau_b) \over {  (S_2-S_1) }} \right)~~~~~,  
\label{etafr5}
\end{equation}                                                                
leading to the expression for the efficiency:
\begin{equation} 
{\eta}_{Rf}~=~~ 
{ \omega_a \over \omega_b } 
 { 1 \over { \eta^{fricles}_{H.E.}~+~ { \sigma^2(1/\tau_a~+~1/\tau_b)
\over {S_2-S_1} } }}~<~       
{ T_c \over T_h } 
{ 1 \over { \eta^{fricles}_{H.E.}~+~ { \sigma^2(1/\tau_a~+~1/\tau_b)
\over {S_2-S_1} } }}                                                
\label{etafr6}                                                                
\end{equation}                  
For both the heat engine and the heat pump, the 
efficiency is explicitly dependent on time allocation, 
cycle time, and bath temperatures.

\subsubsection{Optimization}
\label{optimization}

The performance of both the heat engine and the heat pump can be 
optimized with respect to:
\begin{itemize}  
\item{(a) The overall time period $ \tau$ of the cycle, 
          and its allocation between the hot and cold branches.}

\item{(b) The overall optimal time allocation between all branches.
          (This optimization is performed only for the heat pump.)}
 
\item{(c) The external fields,  ($ \omega_a $, $ \omega_b $).}
\end{itemize}

(a) \bf Optimization with respect to time allocation. \rm

The optimization of time allocation is carried out
with the constant fields $ \omega_a $ and $ \omega_b $. 
The Lagrangian for the work output becomes:
\begin{eqnarray} 
{\cal L}(x,y, \lambda)~=~~{\cal W}_{cycle} +
\lambda
\left(
\tau + {1 \over {\Gamma_c}} \ln (x) +  {1 \over {\Gamma_h}} \ln (y)
- \tau_a - \tau_b
\right)~~, 
\label{langrange1}
\end{eqnarray}                                                                
where $\lambda$ is a Lagrange multiplier.
Equating the partial derivatives of ${\cal L}(x,y, \lambda)$ with respect
to $x$ and $y$ to zero,  the following condition for the optimal
time allocation becomes:
\begin{eqnarray}
\begin{array}{l}
\Gamma_c ~x
\left(
 (1-y)^2 (S_h^{eq}-S_c^{eq}) ~+~\sigma^2 (1-y) (1/\tau_a~+~y/\tau_b)
\right)
~=~\\
\Gamma_h ~y
\left(
 (1-x)^2 (S_h^{eq}-S_c^{eq}) ~-~\sigma^2 (1-x) (x/\tau_a~+~1/\tau_b)
\right)    
\end{array}                                                                   
\label{optimal2}
\end{eqnarray}                                             
                   
When $\sigma=0$, the previous frictionless result is retrieved.
(Optimizing the entropy production $\Delta S^u$ leads to an identical
time allocation to Eq. (\ref{optimal2})).

Eq. (\ref{optimal2}) can also be written in the following way:
\begin{eqnarray}
\Gamma_c ~x \left( (1-y) (1~-~y~x_{max}) \right)
~=~
\Gamma_h ~y \left( (1-x) (x_{max}~-~x) \right)
\label{optimal10f}
\end{eqnarray}  
where $  x_{max} $ was defined in Eq. (\ref{minxc}).
The result is dependent on the time allocations
of the 'adiabats', through the dependence of   $  x_{max} $.
 
For the special case when $ \Gamma_c~=~\Gamma_h $, the relation
between the time allocations in contact with
the hot and cold baths becomes:
\begin{eqnarray}
x~~=~~~x_{max}~y
\label{optimal20f}
\end{eqnarray}
For the frictionless case, 
this result coincides with the former frictionless one$ x=y$,
meaning that equal time is allocated
to contact with  the cold and hot reservoirs. 
When friction is added this symmetry is broken, 
Eq. (\ref{optimal20f}), to compensate for the additional heat 
generated by friction
the time  allocated to the cold  branch, becomes larger than 
the time on the hot branch.

The Lagrangian for the heat-flow, ${\cal Q}_F$, extracted
from the cold reservoir is defined in parallel
to the Lagrangian for the total work.  
Substituting $ \Gamma_h $ for $\Gamma_c $, $x$ for $y$
and vice versa, also $ y_{max} $ for $ x_{max} $,
where  $  y_{max} $ was defined in Eq. (\ref{maxyh}),
one gets the optimal time allocation for the heat pump.

\begin{figure}[tb]
\vspace{-.4cm}
\psfig{figure=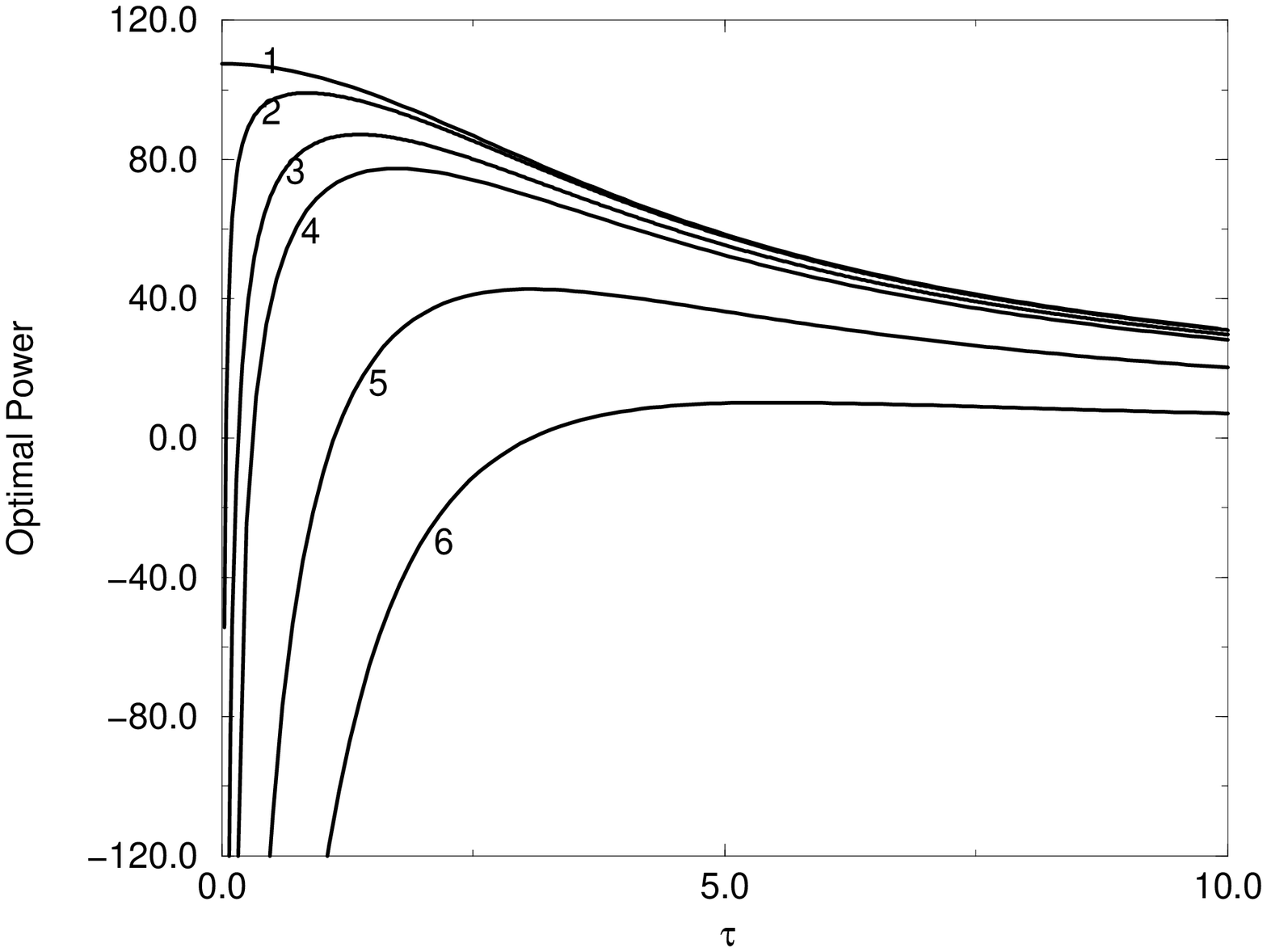,width=0.3\textwidth}
\vspace{-4.1cm}
\hspace{4.8cm}
\psfig{figure=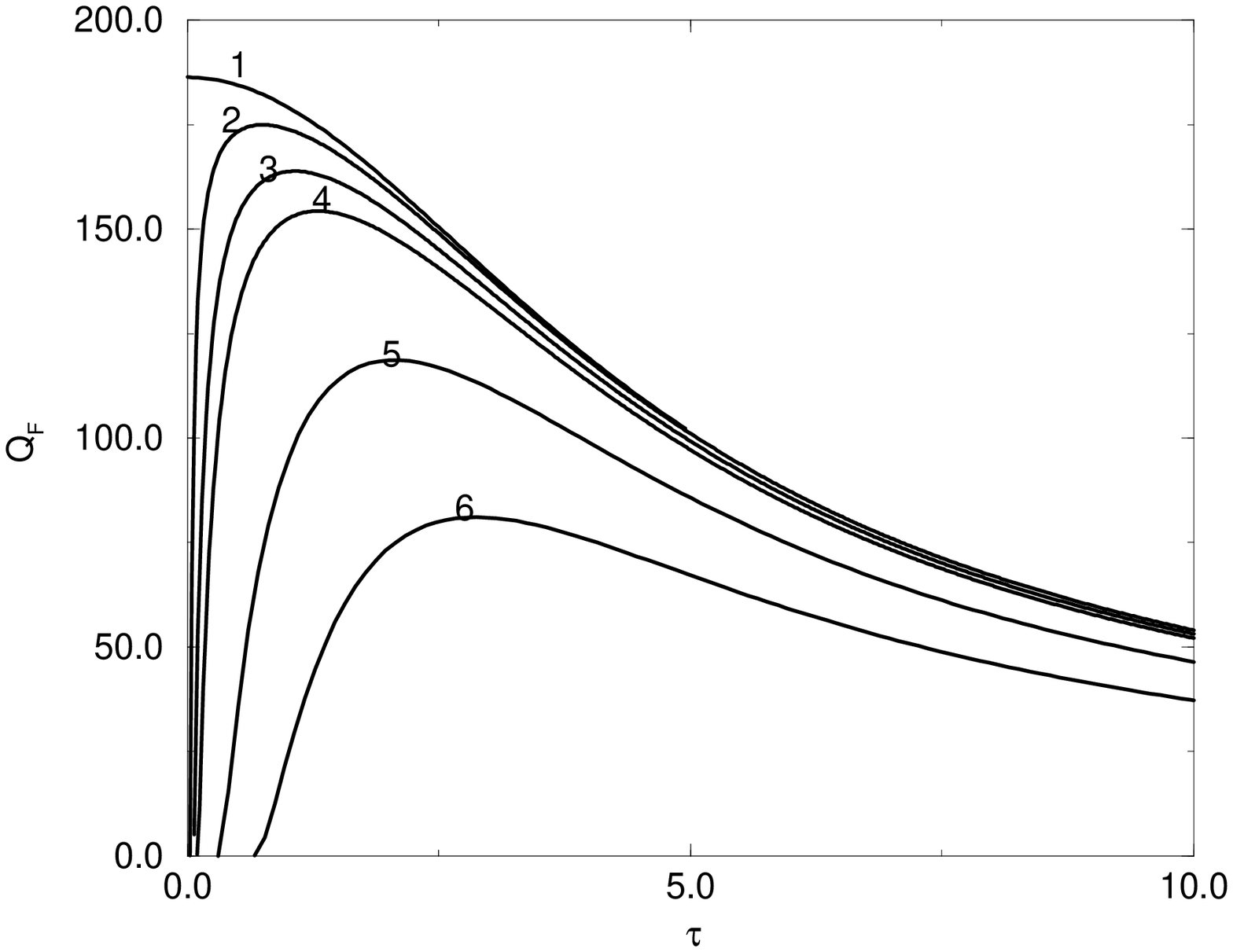,width=0.3\textwidth}
\vspace{-4.1cm}
\hspace{.1cm}
\psfig{figure=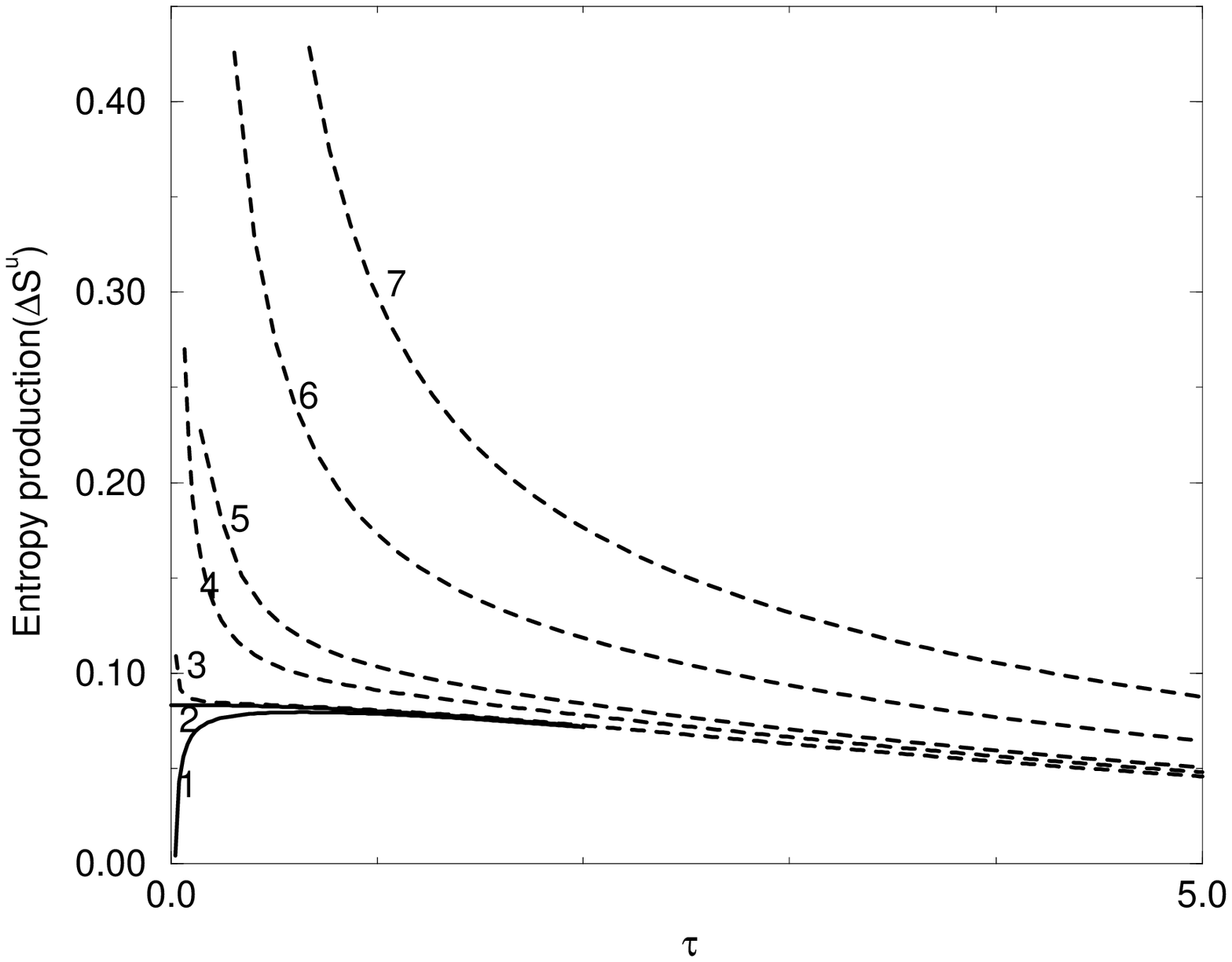,width=0.3\textwidth}
\vspace{4.4cm}
\caption{
Left: The optimal power with respect to time allocation
as a function of $\tau$, for different values of friction.~ 
Middle: The corresponding heat-flow(${\cal Q}_F$). 
The parameter values for both Left and Middle are: 
for plot 1, $ \tau_a$ = $ \tau_b$ =$ \sigma$ = 0.
For all the other plots  $ \tau_a$ = $ \tau_b$ = 0.01
The $ \sigma $ values for the curves from plot 2 to 6 are:
0.002, 0.005, 0.007, 0.0135, 0.02 respectively.
Right: The entropy production rate corresponding to the 
optimal power on
the left part of the figure. 
The additional curve is curve '1', 
which corresponds to $ \sigma$ = 0, 
and $ \tau_a$ = $ \tau_b$ = 0.01 .
The parameter values for the other plots are : 
for plot 2, 
$ \tau_a$ = $ \tau_b$ =$ \sigma$  = 0,
for all the other plots~ $\tau_a$~=~$ \tau_b$~=~0.01.
The $ \sigma$ for the curves from plot~3~to plot~7~are:
0.002, 0.005, 0.007 0.0135, 0.02 accordingly. }
\label{fig:optimp1}
\end{figure}

Optimization of power with respect to time allocation
as a function of  the cycle time, $ \tau$ for different friction coefficients
is shown in Fig.{ \ref{fig:optimp1}}  (Left), together with
the corresponding heat-flow (Middle) and the corresponding
entropy production (Right).  
The left part shows that in the  frictionless case the power 
obtains its maximum 
at zero cycle time with the value
consistent with Eq. (\ref{shorttime}).
When friction is introduced, the maximum power decreases
and is shifted to longer cycle times.
The figure also shows, that for short times the work done 
by  the system is negative, and as the friction coefficient $ \sigma $ 
increases, the boundary between positive and negative power  shifts
to longer cycle times.
In the Middle of Fig. \ref{fig:optimp1},  
the heat-flow corresponding to the 
optimal power on  the left is shown. The shapes of the
power and heat flow curves are similar. 
The heat-flow values are always positive and 
larger than the corresponding power values.
The entropy production (Right) shows that unlike the power curves
the friction changes 
significantly the shape of the curves.
The entropy production rate for the case with friction 
sharply decreases. 
The parallel graphs for the heat pump are similar. 

\bf (b) Time allocation optimization between all branches of 
the refrigerator  \rm

Further optimization of the performance of the heat pump 
is possible by relaxing the assumption of constant time on the 'adiabats'.
First the time allocation between the two 'adiabats'  is optimized, when  
$\tau_a~+~\tau_b~=\delta$, where $\delta$ is a constant.  
Finally the time allocation between the  'adiabats' and the heat exchange 
branches,
is optimized. These results are 
compared to  the recent analysis of Gordon et. all. \cite{gordon97}.

From Eqs. (\ref{rheatf0}) and  (\ref{eqS2S1})
with constant time allocations along the heat exchange branches
one gets for the cooling power:
\begin{eqnarray}
~~~{\cal Q}_F~~~= A_0~-~A_1( {y \over \tau_a} ~+~{ 1 \over (\delta-\tau_a)}) ~~~~,
\label{rheatfj}
\end{eqnarray}
where $A_0$ and $A_1$ are constant functions of  the parameters
of the system. And on $\delta$, a double inequality is imposed
$ \tau~ > \delta ~ > $ ~the larger of [ $(\tau~-~\tau_{h,min}); \tau_{b,min}$],
see Eq. (\ref{tbmin}).

The optimal $\tau_a$ depends only on $y$ and
on $\delta$. The optimal value of $\tau_{a,opt}$ becomes:
\begin{eqnarray}
~~~\tau_{a,opt}~~=~~ \delta { -y~+~ \sqrt{y} \over
1-y } 
\label{tauaop}
\end{eqnarray}

Further optimization by changing the the value of $\delta$, 
changes the cycle time $\tau$.
This  optimization step is done by numerical iteration.
Typically  the sum of the final optimal values of 
$\tau_a$~and $\tau_b$ is about
twice their value before, and their ratio is about 0.7 of 
the value which was chosen initially. 

The next step is to study the time allocation between the 
'adiabats' and the heat exchange branches when all other
controls of the heat pump have optimal values.
These controls include also the external fields of 
optimization which are described later.

For comparison with Gordon et. all. \cite{gordon97}, 
the results of optimization
are plotted  in the $ 1/{\cal Q}_F$ ,$1/\eta$ plane for a fixed cycle time
$\tau$. The following example demonstrates the method followed:
First an optimal starting value for ${\cal Q}_F$ was found which
determines the time allocation control parameters,
$ ~\tau_c=0.44221,~~ \tau_h=0.31779,~~ \tau_a=0.0084,~~ \tau_b=0.0116$ 
with a total cycle time of $\tau=0.78$.
Under such conditions ${\cal Q}_{F,max}=2.9158$ ($1/{\cal Q}_{F,max}=0.34296$). 

\begin{figure}[tb]
\vspace{1.2cm}
\psfig{figure=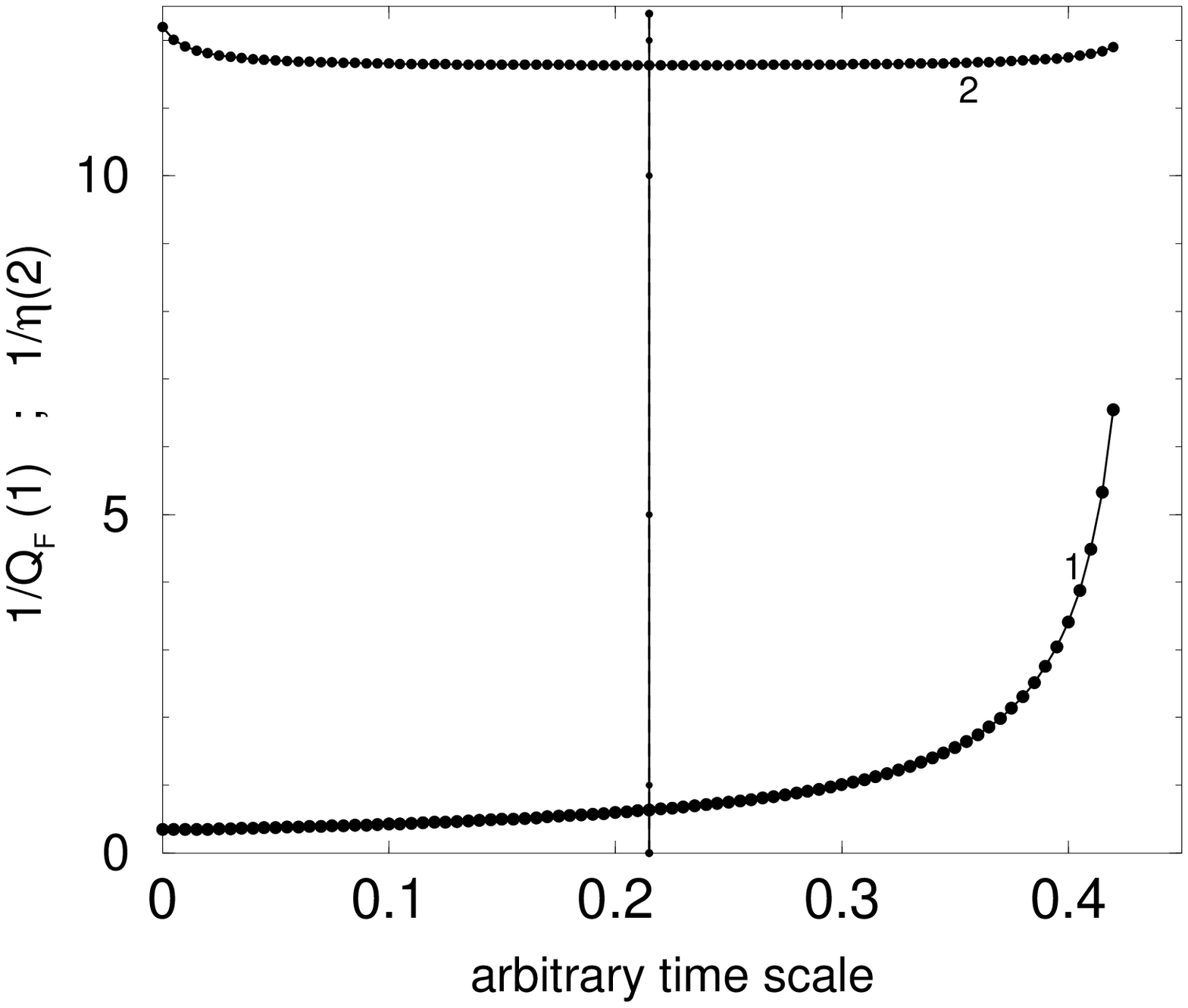,width=0.43\textwidth}
\vspace{-6.78cm}
\hspace{7.cm}
\psfig{figure=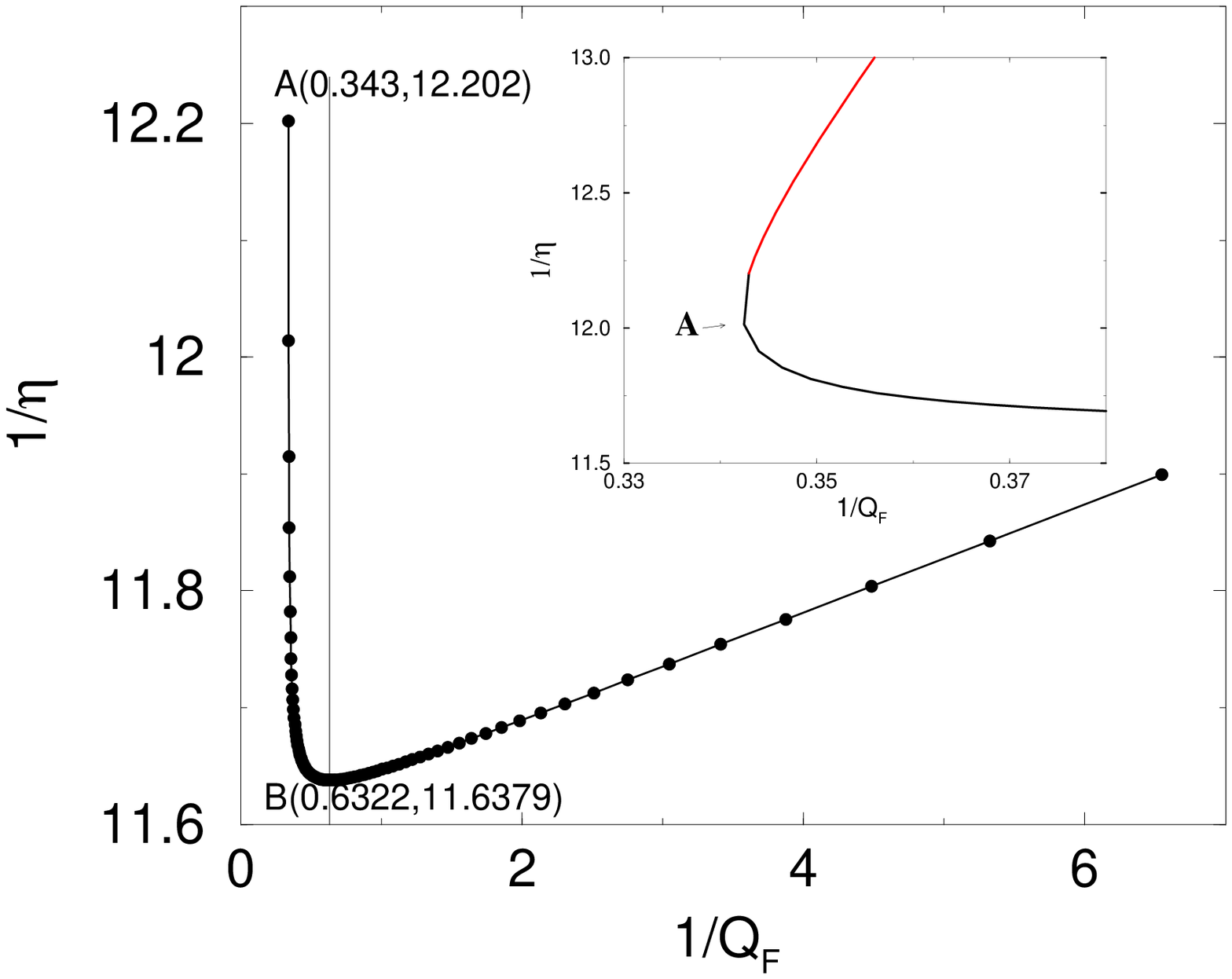,width=0.43\textwidth}
\vspace{0.1cm}
\caption{The relation between efficiency and cooling power for the heat pump.
The parameters are: 
The constant optimal cycle time, $\tau$,~=0.78;
T$_c$=51.49, T$_h$=257.45, $\omega_a$=47.699, $\omega_b$=600,
$\Gamma_c$=1, $\Gamma_h$=2, $\sigma$=0.005  
Left: Comparison between $1/{\cal Q}_F$(plot 1) and $1/\eta$(plot 2) 
as a function of the allocated time transfer from the heat exchange branches 
to the 'adiabats'. Zero time is the optimal heat-flow time allocation.
Right:  The Universal plot for the heat pump.
The starting optimal  point in the plane of ($1/{\cal Q}_F$, $1/\eta$),
was (0.34296,~12.202), while the maximum efficiency point $\bf B$ is
(0.6322,11.6379)
and time allocation $ (\tau_c, \tau_h, \tau_a, \tau_b)$=
(0.22721,0.16328,0.1636,0.2259).
 The insert shows the neighborhood of point $\bf A$.}
\label{fig:univp1}
\end{figure}

Changing the time allocation between the 'adiabats' and the heat exchange 
branches changes the balance between optimal cooling power and efficiency.
Denoting the sum $ \tau_c $+$ \tau_h $ 
by $ \tau_{ch} $, the ratio $\tau_h/\tau_c$ by $r_{hc}$, the
sum $\tau_a$~+~$\tau_b$ by $\tau_{ab}$,
the ratio $\tau_a/\tau_b$ by $r_{ab}$,  
time is transfered from  $\tau_{ch}$ by small steps to  $\tau_{ab}$,
while keeping the the ratios $r_{hc}$ and $r_{ab}$ constant.
For each step the corresponding 
$ 1/{\cal Q}_F$ and $1/\eta$, are calculated as in Fig.{ \ref{fig:univp1}}.
The relation between the reciprocal efficiency and the 
reciprocal cooling power shows the tradeoff between losses 
due to friction and losses
due to heat transfer. Following the curve in Fig. \ref{fig:univp1}, 
starting from point $\bf A$ where the cooling power is optimal,
resources represented by time allocation are transferred 
from the heat exchange branches
to the 'adiabats', reducing the friction losses. 
At point $\bf B$ an optimum is reached
for the efficiency. This point has been found by Gordon et. al. 
to be the universal operating choice for commercial chillers.
Point $\bf B$ represents the optimal compromise 
between maximum efficiency and cooling power.
Point $\bf A$ is located at the maximum  cooling power. 
If more time is allocated to the heat exchange branches both
$1/{\cal Q}_F$ and $1/\eta$ will continue to increase as seen
in the insert of Fig. \ref{fig:univp1}.

(c) \bf   Optimization with respect to the fields. \rm 

The values of the fields $\omega_a$ and $\omega_b$ are control parameters of the 
engine. In a spin system these fields are equivalent to the value of the
external magnetic field applied on the system. They directly influence
the energy spacing of the TLS.
The work function ${\cal W}_{cycle}$, or equivalently the power ($\cal P$)
is optimized with respect to the fields, subject to the Carnot constraint:
\begin{equation}
{\omega_a \over T_c }~ \geq~ { \omega_b \over T_h }
\label{langrange2}
\end{equation}
Optimal power is obtained by                                                  
equating independently to zero  the partial derivatives of 
$~~{\cal W}_{cycle}$, or of ${\cal P}~=~~{\cal W}_{cycle}/\tau $ 
by varying $\omega_a$ and $\omega_b$.
In addition the optimal solutions have to fulfill the inequality 
constraints in  Eq. (\ref{langrange2}). As a result two transcendental
equations in  $\omega_a$ and $\omega_b$ are obtained
which are solved numerically.

The two equations are:
\begin{eqnarray}
\begin{array}{r}
~{ (1~-~y~x_{max}) \over (\omega_b~-~\omega_a)}
~({ \Delta S^{eq}~+~ \sigma^2/\tau_a})
~  \cosh^2 \left( {{\omega_a} \over {2 k_B T_c} } \right)~=~
{~1-y \over (4~k_B~T_c) }
\\
~{ (x_{max}~-~x) \over (\omega_b~-~\omega_a)}
~({ \Delta S^{eq}~+~ \sigma^2/\tau_a})
~  \cosh^2 \left( {{\omega_b} \over {2 k_B T_h} } \right)~=~
{~1-x \over (4~k_B~T_h) }
\end{array}
\label{optimal4}
\end{eqnarray}                                                                

Where $ \Delta S^{eq}$= S$_h^{eq}$- S$_c^{eq}$ as  defined in 
Eq. (\ref{eqS1S2}).
Examining Eq. (\ref{optimal4}), 
and fixing the friction $ \sigma$, it is found that $ \Delta S^{eq}$
is an extensive function of order zero (intensive ) with respect to the
quartet of variables $ \omega_a, T_c, \omega_b, T_h$. 
This means that scaling these parameters simultaneously will
not change $ \Delta S^{eq}$.  
Also x$_{max}$,
and  $ \cosh^2 \left( {{\omega} \over {2 k_B T}} \right)$
are extensive (order zero).
The work function however, is extensive with order one
(Eqs. (\ref{WcycH2}) and  (\ref{WcycH3})).
This property will be exploited in paragraph III.

\begin{figure}[tb]
\vspace{-1.2cm}
\hspace{1.5cm}
\psfig{figure=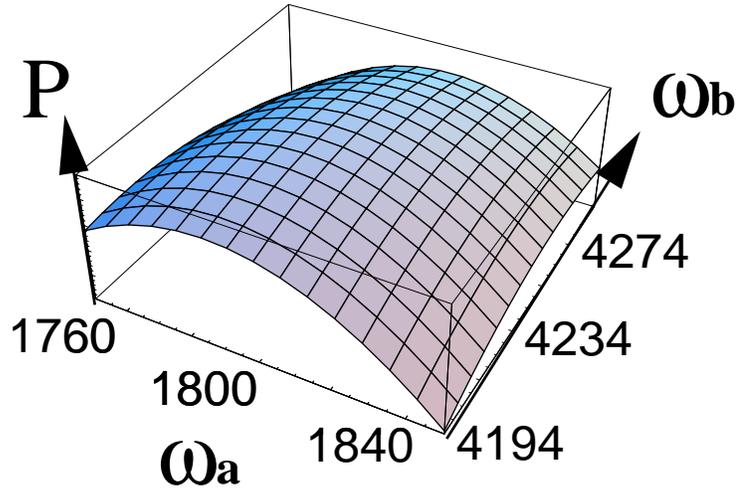,width=0.7\textwidth}
\vspace{0.1cm}
\caption{Power for the frictionless engine as a function of the fields
($ \omega_a $, $ \omega_b $) 
for constant bath temperatures, and  constant  time allocations.
The maximum power is achieved at
$\omega_a=1794 $ and $\omega_b=4239$ where the bath temperatures are
 $T_c=500, T_h=2500$. }
\label{fig:fig4}
\end{figure}

The optimization of power with respect 
to the fields is shown in 
Fig. \ref{fig:fig4} for the frictionless engine, 
as a function of the fields with 
fixed time allocation. A global maximum can be identified.

The heat pump optimization of ${\cal Q}_F$ with respect to the fields is different 
and therefore will be presented in Section \ref{rrof}.

The analysis for the optimization with respect to the fields for the
entropy production $\Delta {\cal S}^u$, is presented in appendix \ref{sec:entrfricA}.
The optimal solution without friction($\sigma=0$) leads to
 $\Delta {\cal S}^u_{min}=0$. When 
$\sigma \neq 0$, the minimum value of
$\Delta {\cal S}^u$ is different from zero, and is achieved on the 
boundary of the region.

\subsection{Global Optimization of the Heat Engine} 
\label{sec:globalopt}

Global optimization of the power means searching for the optimimum
with respect to the control parameters
cycle time, time  allocation  and the fields.
An iterative procedure is used.

The procedure is initiated by setting 
the optimal time allocation from the corresponding Lagrangian, 
with $\sigma = 0$. The power becomes  
a product of two functions, one depending only on time
the other only on the fields, and therefore,  
the fields can be changed independently of time. The optimal
fields for the  above time allocation are then sought.
For the frictionless case, the overall time on the adiabats tends to zero. 
The optimal field values become independent of
time. The value ${\cal P}=107.501$ is the short time limit in accordance 
with the equation:
\begin{eqnarray}
{\cal P}~~\longrightarrow~~( \omega_b - \omega_a )(S_{h}^{eq}-S_{c}^{eq})
(\Gamma_{c} \Gamma_{h})/(\sqrt{\Gamma_c}+\sqrt{\Gamma_h})^2
\label{shorttime}
\end{eqnarray}

These fields are inserted into the expression with 
friction $\sigma \neq 0$, 
and the new optimal times and fields are computed.
The iteration converges after two to three steps, 
as indicated by  Table \ref{tab:iterat2} for $ \sigma=0.005$.
Notice that the location of the optimum is not very sensitive 
to the friction parameter.
\newpage
\begin{table}
\caption{Global optimization of power. 
The notation ${\cal P}^{max}(\omega_a,\omega_b)$ stands for fixed time allocations,
and the notation ${\cal P}^{max}(\tau_a,\tau_b)$ stands for fixed field values.
The other parameters are: $T_c=500$, $T_h=2500$, $ \tau_a$= $ \tau_b$=0.01.  
$\Gamma_c=1$ and $\Gamma_h=2$}
\begin{center}
\begin{tabular}{||c|c|c|c|c|c|c||}
\tableline
$ \sigma$ & $ \tau$ &  ${\cal P}^{max}( \omega_a, \omega_b)$ &     
 ${\cal P}^{max}( \tau_c, \tau_h)$ & $ \omega_a$ &  $ \omega_b$ & 
$ \tau_c/  \tau $  \\
\tableline
0.005  & 2 &  & 84.46  & 1794  & 4239 & 0.5999 \\
\tableline
0.005  & 1.367 &  & 87.18  & 1794  & 4239 & 0.5891 \\
\tableline
0.005  & 1.367 & 88.68 &  & 1719.1  & 4036.31 & 0.5891 \\
\tableline
0.005  & 1.347 &  & 87.47  & 1719.1  & 4036.31 & 0.58856 \\
\tableline
0.005  & 1.347 & 88.704  &  & 1718.16  & 4033.67 & 0.58856 \\     
\tableline
\end{tabular}
\end{center}
\label{tab:iterat2}
\vspace{0.4cm}
In Table \ref{tab:iterat3}, the extensive properties  
Eq.  (\ref{optimal4}) are examined 
for k=2 and k=10 with respect to Table \ref{tab:iterat2}. 
The temperature values will change to T$_c~=~1000$,
T$_h~=~5000$ for k=2 and  T$_c~=~5000$,
T$_h~=~25000$ for k=10~. The results verify the analysis.
\vspace{0.4cm}
\caption{Global optimization of power. By multiplying the
four values, T$_c$,~T$_h$, $ \omega_a$, $ \omega_b$ by k 
and searching first for optimal time allocation, then multiplying
only the temperature values by k and searching for the optimal
fields. All the notations and other parameters as in
 Table \ref{tab:iterat2}.    }
\begin{center}
\begin{tabular}{||c|c|c|c|c|c|c|c||}
\tableline
$ \sigma$ & k & $ \tau$ &  ${\cal P}^{max}( \omega_a, \omega_b)$ &     
 ${\cal P}^{max}( \tau_c, \tau_h)$ & $ \omega_a$ &  $ \omega_b$ & 
$ \tau_c/  \tau $  \\
\tableline
0.005  & 2 & 1.367  &   & 174.9  & 3438.2  & 8072.6 & 0.58852 \\
\tableline
0.005  & 2 & 1.367    & 174.9 &  & 3436.7  & 8070.3 & 0.58852 \\
\tableline
0.005  & 2 & 1.347  & & 174.94 & 3436.7  & 8070.3 & 0.58856 \\
\tableline
0.005  & 2 & 1.347 & 179.8 &   & 3437.7  & 8069.4 & 0.58856 \\
\tableline
0.005  & 10 & 1.347 & 887.04 &  & 17181.6  & 40336.7 & 0.58856 \\     
\tableline
\end{tabular}
\end{center}
\label{tab:iterat3}
\end{table}

\section{Asymptotic Properties of the Heat Pump when the cold bath 
temperature approaches absolute zero.}
\label{rrof}
The goal is to obtain an asymptotic upper bound on the cooling power
when the heat pump is operating close to absolute zero temperature.
This requires optimizing the performance of the heat pump with respect to all
control parameters.

\subsection{Optimization of the heat-flow ${\cal Q}_F$ with respect 
to the fields and to the cooling power upper bound.}

The heat-flow,${\cal Q}_F$ extracted from the cold 
reservoir now becomes the subject of interest:
\begin{eqnarray}
~~~{\cal Q}_F~~~= \omega_a (S_2-S_1)/ \tau 
\label{rheatf1}
\end{eqnarray}
or from  Eq. (\ref{eqS2S1}),
\begin{eqnarray}
~~~{\cal Q}_F~~~= ( \omega_a/\tau ) \left(
  ( S_2^{eq}- S_1^{eq}) F(x,y) ~-~  
        {  \sigma^2  (1-x)  (y/\tau_a + 1/\tau_b) 
            \over (1-xy)  } \right)
\label{rheatf2}
\end{eqnarray}

No global maximum for the  ${\cal Q}_F$ with respect to the fields is found.
The derivative of ${\cal Q}_F$ with respect to  $ \omega_b$ becomes: 
\begin{eqnarray}
{ {\partial {{\cal Q}_F}} \over { {\partial {\omega_b} } }~}=
~~{~F(x,y)~\omega_a \over \tau~}~ { 1 \over 4~k_B~T_h~\cosh^2{ \omega_b
\over 2~k_B~T_h } }  \geq 0
\label{rheatf5}
\end{eqnarray}
leading to the result  that ${\cal Q}_F$ is monotonic in $ \omega_b$.
Under such conditions, $ \omega_b$ is set, and the optimum with respect
to  $ \omega_a$ is sought for.
The derivative of ${\cal Q}_F$ with respect to $ \omega_a$ becomes: 
\begin{eqnarray}
{ {\partial {{\cal Q}_F}} \over { {\partial {\omega_a} } } }~=
( S_2^{eq}- S_1^{eq})  ~-~  
{  \sigma^2 \over  (1-y) } (y/\tau_a + 1/\tau_b)~-~ \omega_a~
 {1 \over 4~k_B~T_c~ \cosh^2{\omega_a \over 2~k_B~T_c}}~=~0
\label{rheatf6}
\end{eqnarray}
Introducing from  Eq. (\ref{rheatf6}) the optimal value of
$( S_2^{eq}- S_1^{eq})  ~-~  
{  \sigma^2 \over  (1-y) } (y/\tau_a + 1/\tau_b)$,
into   Eq. (\ref{rheatf2}),
leads to the optimal cooling rate:
\begin{eqnarray}
~~~{\cal Q}_F^{optimum}~~~=
~~{~F(x,y)~\omega_a^2 \over \tau~}~ { 1 \over 4~k_B~T_c~
\cosh^2{ \omega_a
\over 2~k_B~T_c } }~=~
~{~F(x,y)~ \over  4~k_B~ \tau~}~
\left( { ~\omega_a  \over ~T_c~ } \right)^2~{ T_c \over
 \cosh^2{ \omega_a
\over 2~k_B~T_c } }
\label{rheatf21}
\end{eqnarray}

Due to  its extensivity, the ratio
${ \omega_a \over T_c } $ becomes a constant,
while both $ \omega_a$ and T$_c$ can approach zero.

From Eq.(\ref{rheatf21}), an upper-bound for the cooling 
rate ${\cal Q}_F$ is obtained:
\begin{eqnarray}
~~~{\cal Q}_F^{optimum}~~~ \leq~~
~{~F(x,y)~ \over  4~k_B~ \tau~}~
\left( { ~\omega_a  \over ~T_c~ } \right)^2~T_c     . 
\label{rheatf22}
\end{eqnarray}
From Eqs. (\ref{rheatf22}), when T$_c$ approaches zero,
the cooling rate vanishes, at least linearly with temperature.
This is a third law statement which shows that 
absolute zero cannot be reached since the
rate of cooling vanishes as absolute zero is approached. 

\subsection{The asymptotic relation between the internal and 
external temperature
on the cold branch}

When  the bath temperature tends to zero, the 
internal working fluid temperature has to follow.
This becomes a linear relationship  between $T'$ and $T_c$
as $T_c$ tends to zero.  

Calculating the polarization at the end of the contact with the cold bath
$S_2$:
\begin{eqnarray}
S_2= S_2^{eq}~-~{ (S_2^{eq}-S_1^{eq}) x (1-y)~-~ \sigma^2 x
 (1/\tau_b~+y/\tau_a)  \over (1-xy)  }
\label{eqS1r}
\end{eqnarray}
Assuming the relation $T_h$~=~$\rho$~$T_c$ as $T_c$ tends to zero,
the exponents can be expanded to the first order to give:
\begin{eqnarray}
{S_2}=~{ T_c \over \omega_a }
{ 1 - x y_{max} + (\omega_a/\omega_b) \rho  x(y_{max}-y)
 \over (1-xy)  }~~
+~~ 1/2~-~{ x (\sigma^2/\tau_a)(y_{max}-y)  \over (1-xy)  }
\label{eqS2r}
\end{eqnarray}
Also, $S_2$ defines the internal temperature $T'$ through the relation:
${S_2}~ = ~-{1  \over 2} \tanh \left( {{\omega}_a 
\over {2 k_B T'}} \right)~~$.
Expanding the hyperbolic tangent, one gets:
\begin{eqnarray}
{T'}~~=~~
 {T_c}~{ 1~-~x y_{max}~+~ \rho (\omega_a/\omega_b) x (y_{max}-y) 
 \over (1-xy)  }~~
-~~{ x \omega_a( \sigma^2/\tau_a) (y_{max}-y)
  \over (1-xy)  }
\label{eqS4r}
\end{eqnarray}
proving that $T_c$ and $T'$ both tend
asymptotically to zero. It should be noted that the term independent
of $T_c$ depends on $\omega_a$, which also  
tends to zero as $T_c$ tends to zero ( (Eq. \ref{rheatf22}).
Eq. (\ref{rheatf22}) also
shows that ${\cal Q}_F^{optimum}$*T$_c$ is a quadratic function
of $ \omega_a$, Cf. Fig. \ref{fig:qftcwa1}.

Eq.  (\ref{rheatf22}) represents an upper-bound to the rate of cooling. 
In order to determine how closely this limit be approached, a
strategy of cooling must be devised, which re-optimizes 
the cooling power during the changing conditions  
when T$_c$ approaches zero.  

\subsection{Optimal cooling strategy}

The goal is to follow an optimal cooling strategy, which
exploits the properties of the equations and achieves the upper-bound
for the rate of cooling, ${\cal Q}_F$.

The properties of the equations employed are;
\begin{itemize}
\item{i:   The derivative with respect to $\omega_a$ of ${\cal Q}_F$  
( Eq. (\ref{rheatf6})), is extensive of order zero
in the 'quartet' ($\omega_a, \omega_b, T_c, T_h$).}
 
\item{ii:  For  $ {\partial {{\cal Q}_F}} \over { {\partial {\omega_a} } } $ 
the extensivity holds also for  
the 'doublets' ($\omega_b~,~T_h$) or
($\omega_a~,~T_c$).   Scaling these variables by the same number,
leaves  Eq. (\ref{rheatf6}) equal to zero, and the value of
${\cal Q}_F^{optimum}$ does not change.}

\item{iii: In spite of ${\cal Q}_F$ being monotonic in $ \omega_b$,
$~{\cal Q}_F^{optimum}$ is independent of $ \omega_b$ (and of T$_h$),
therefore ${\cal Q}_F$  saturates  as  $ \omega_b$ is increased.}
\end{itemize}

From property (i) it follows, that once an optimal
'quartet'($ \omega_a, \omega_b, T_c, T_h$) is created, it is possible to
cool optimally with a set of quartets, which are scaled
by a decreasing set {$r_n~<~1$}, $\lim_{n\rightarrow \infty} r_n =0$. 
For this set the limit of the ratio
${ \omega_a \over T_c } $
is a non zero constant. Therefore in Eq. (\ref{rheatf21})  
$\omega_a$ and T$_C$ are optimal  leading to:
\begin{eqnarray}
~~~{\cal Q}_F^{optimum}~~~=
~{~F(x,y)~ \over  4~k_B~ \tau_{optimal}~}~
\left( { ~\omega_{a,optimal}  \over ~T_{c,optimal} } \right)^2~
{ T_{c,optimal} \over
 \cosh^2{ \omega_{a,optimal}
\over 2~k_B~T_{c,optimal} } }
\label{rheatf25}
\end{eqnarray}

In general, the hot bath temperature is constant, and the 
property (ii) is used to scale back the value of the optimal T$_h$ 
to the bath temperature. As a result,
 the optimal high field is also scaled.

\begin{figure}[tb]
\vspace{-.5cm}
\psfig{figure=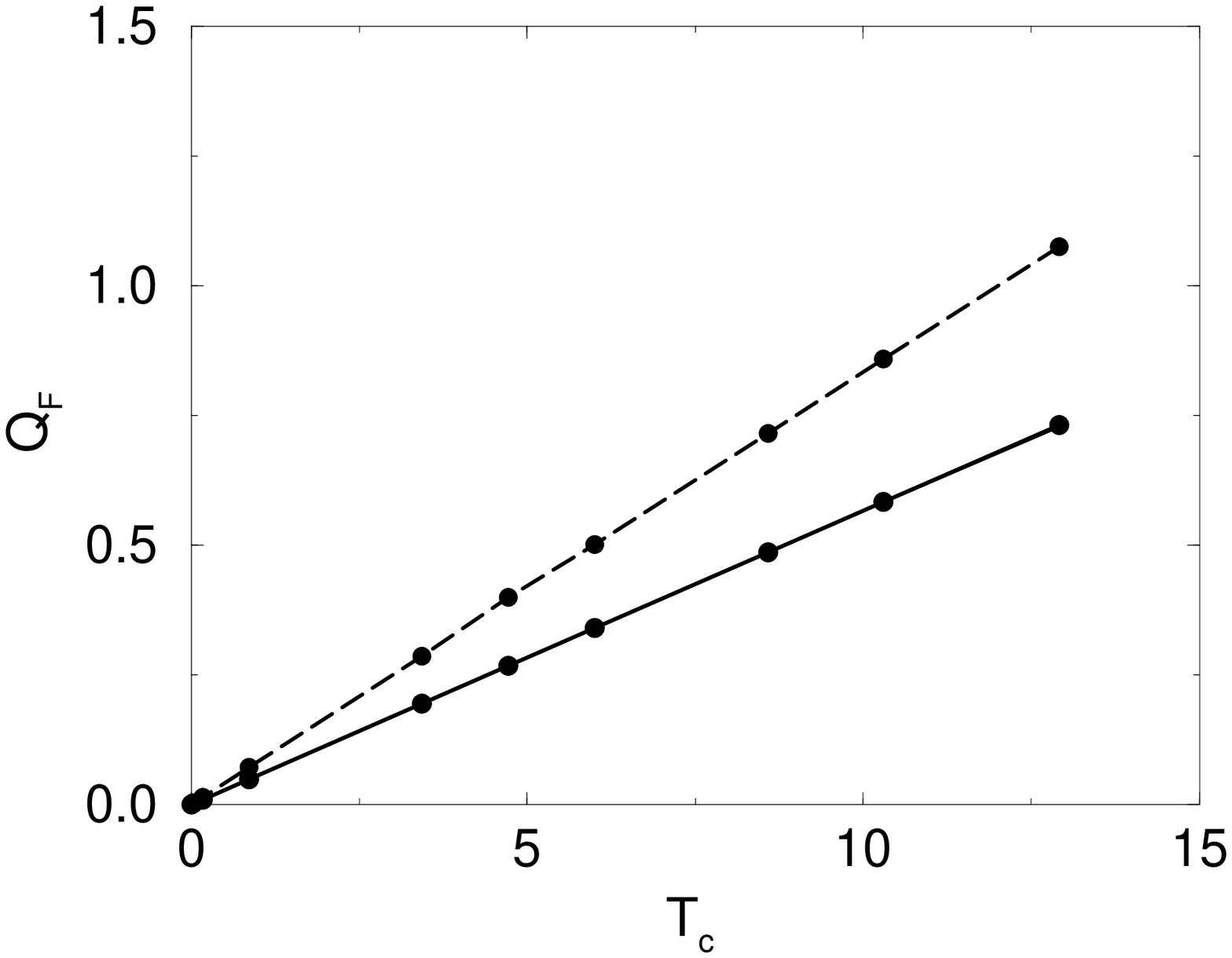,width=0.5\textwidth}
\vspace{-6.78cm}
\hspace{7.cm}
\psfig{figure=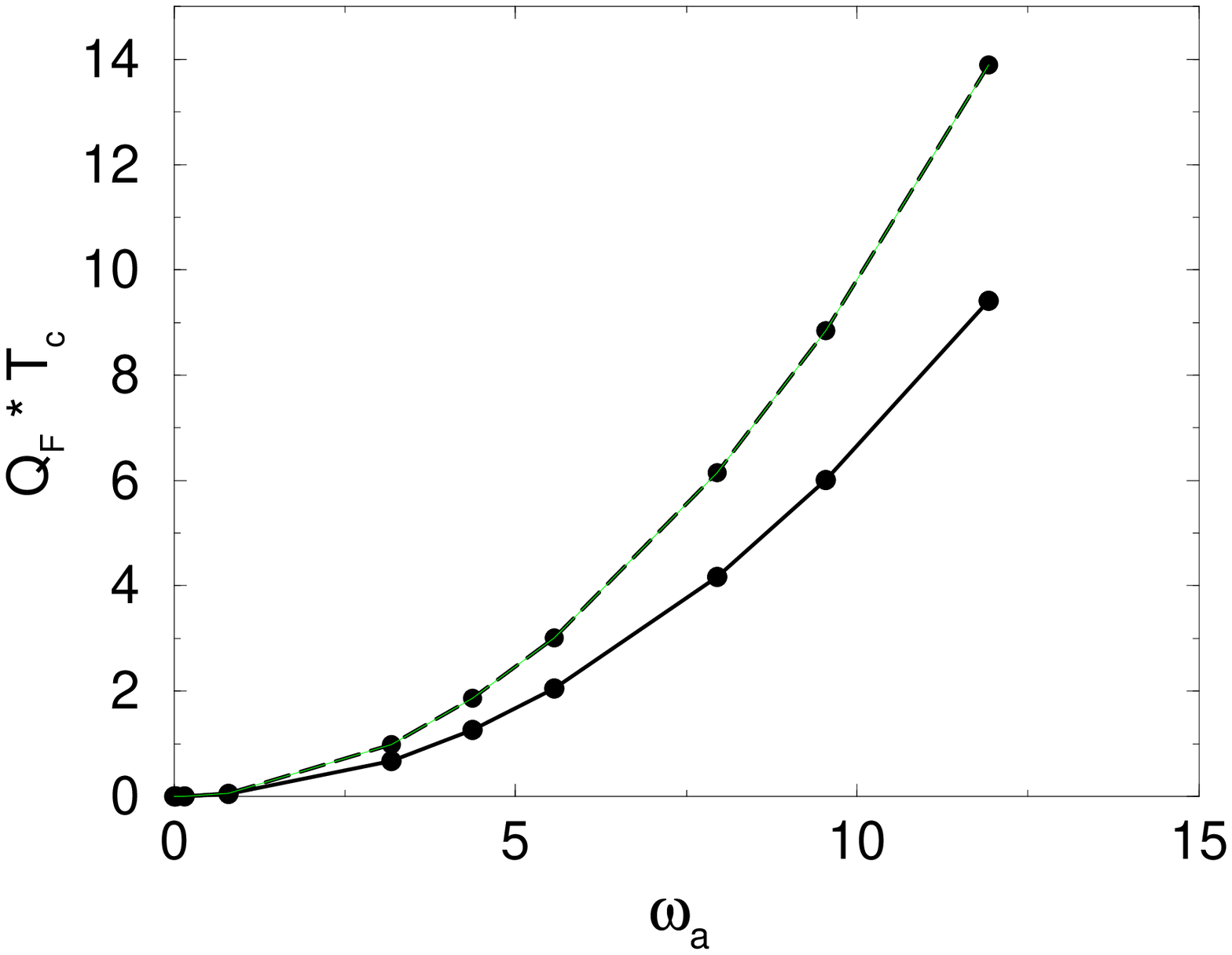,width=0.5\textwidth}
\vspace{0.1cm}
\caption{Left, (solid line): The optimal heat-flow ${\cal Q}_F^{optimum}$
for the heat pump as a function of T$_c$.
The fixed parameter values of the pair ($\omega_b$ and T$_h$) are,
T$_h=64.359, 51.49, 42.9082, 30.03555, 23.599, 17.153,      
4.291, 0.8582, 0.1717, 0.01717$, and
accordingly $ \omega_b=150, 120, 100, 70, 55, 40, 10, 2, 0.4, 0.04$.
The other constant parameter values are:
 $ \sigma=0.005$, $ \Gamma_c=1$, 
$ \Gamma_h=2$. (dashed line): fixing 
$\omega_b~=~3000$ for every point. The other parameters are
the same as the solid line.
Right: The optimal heat-flows multiplied by the corresponding
T$_c$ as a function of $ \omega_a$.
The optimal time is constant for the chosen
parameters; $\tau_{optimal}=0.885$ for the solid curves, and 
0.825 for the dashed curves.   }
\label{fig:qftcwa1}
\end{figure}

Property (iii) will be exploited by changing 
only $ \omega_b$ in  the optimal quartets and checking for saturation.
See Fig.   \ref{fig:qftcwamax}  and the dashed curves of
Fig.  \ref{fig:qftcwa1}.
Summarizing, for every 'quartet' the upper-bound in
Eq.  (\ref{rheatf22}) can be reached. The details of the 
cooling strategy can be found in Appendix \ref{strategy}

\begin{figure}[tb]
\vspace{-.5cm}
\psfig{figure=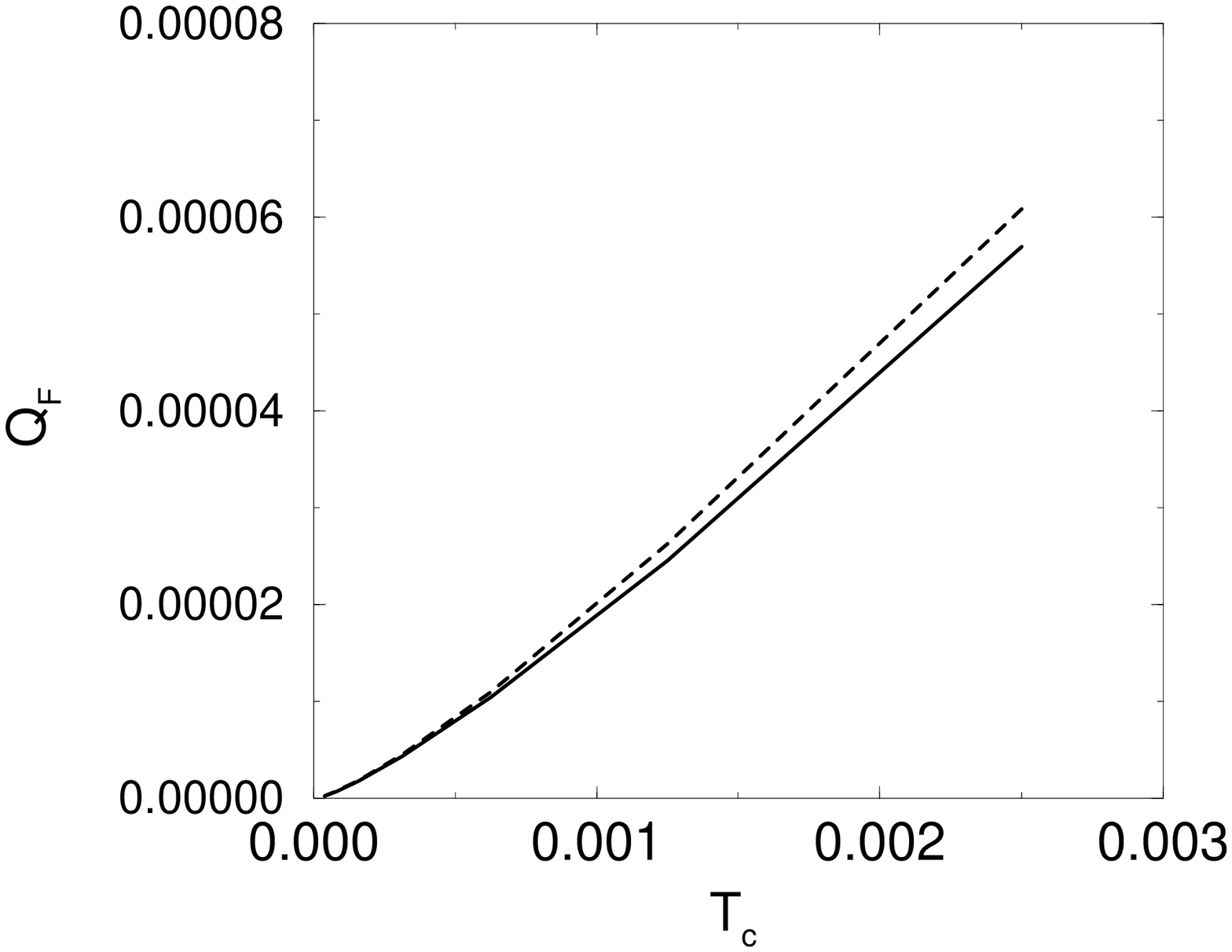,width=0.5\textwidth}
\vspace{-6.78cm}
\hspace{7.cm}
\psfig{figure=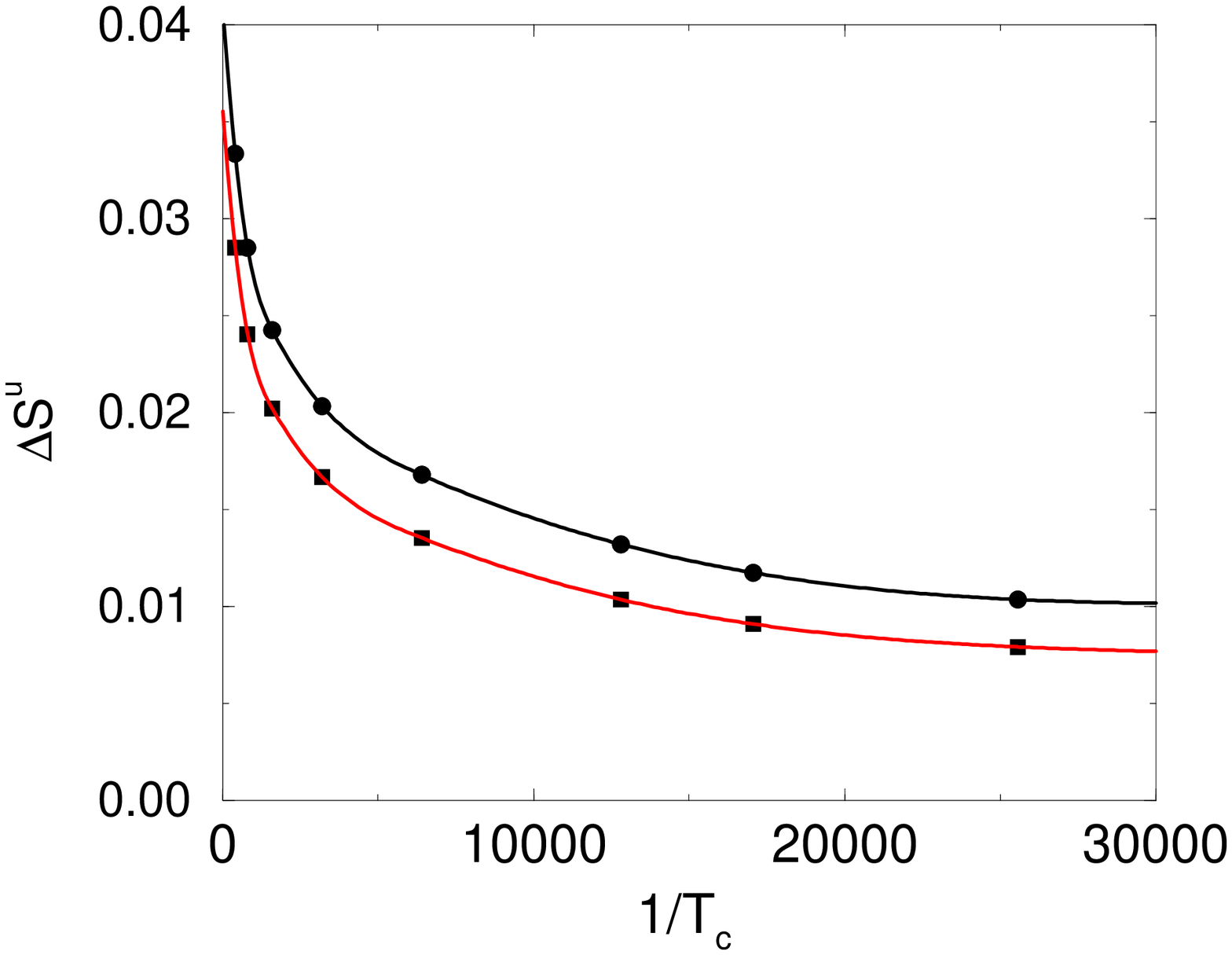,width=0.5\textwidth}
\vspace{0.1cm}
\caption{Left: The optimal cooling rate as a function of the cold bath temperature
T$_c$, compared with  the upper-bound for the cooling rate.
Right: The entropy production during cooling shown for the 
case with friction (upper line, circles) and without friction (lower line, squares).
The common parameters for all three figures are:
$ \tau_a = \tau_b =0.01$, $ \sigma=0.005$, $ \Gamma_c=1$, 
$ \Gamma_h=2$.    }
\label{fig:asympent1}
\end{figure}

Fig. \ref{fig:asympent1} shows that the cooling strategy 
( Tables \ref{tab:zeroset} and  \ref{tab:zeroset2} )
can approach the upper bound leading to a linear relation
of the optimal cooling power with temperature.
With respect to the fields the optimal  strategy leads to 
a decrease of the field $\omega_a$ which is in contact with the cold bath. 
This causes the internal temperature of the TLS T' to be lower than 
the cold bath temperature T$_c$. On the hot side the optimal solution
requires as large an energy separation as possible $\omega_a \rightarrow \infty$
but this effect saturates.

The linear relation of the cooling rate with T$_c$ leads to a 
constant asymptotic
entropy production as can be seen 
in the right of Fig. \ref{fig:asympent1} ( Cf.  Appendix C).

\section{Conclusion}
\label{ref:conclusion}

The detailed study of the  four stroke discrete heat engine with 
internal friction 
serves as a source of insight on the performance of 
refrigerators at temperatures 
which are very close to absolute zero.
The next step is to find out if the  behavior of 
the specific heat pump described in the study can be generalized.
A comparison with other systems studied indicates that the conclusions 
drawn from the model
are generic. As a heat engine the model shows the generic behavior of
maximum power as a function of control parameters found in 
finite time thermodynamics \cite{salamon77,salamon80,andresen83,bejan96}.
This is despite the fact that the heat transfer laws in the microscopic model
of the working fluid are different from the macroscopic laws
such as the Newtonian heat transfer law \cite{geva0}.
When operated as a heat pump with friction,
the present model shows the universal behavior observed 
for commercial chillers \cite{gordon97} caused by a tradeoff between 
allocating resources to the 'adiabats' or to the heat exchange branches. 

Another question is whether
the linear scaling of the optimal cooling power at low cold bath temperatures 
is a universal phenomenon.
For low temperatures the results of the present model can be extended
to a working fluid consisting of an ensemble of harmonic oscillators
or any N-level systems. This is because at the limit of absolute zero
only the two lowest energy levels are relevant. 
When examined, other models with different operating cycles show an 
identical 
behavior. 
For example the
continuous model of a quantum heat engine \cite{geva2}
based on reversing the operation of a laser shows this linear
scaling phenomena.
Another example is  the Ericsson refrigeration cycle Cf. Eq. (23) in the study 
of Chen et al \cite{chen98} which shows the same asymptotic 
linear relationship.

A point of concern is the dependence of 
the heat transfer laws on temperature when absolute zero is approached.
The kinetic parameters $k_\downarrow $ and  $k_\uparrow$ represent
an individual coupling of the two-level-system to the bath.
Considering coefficients derived from gas phase collisions they settle to  a constant 
asymptotic value as the temperature is lowered \cite{forrey}.
The reason is that the slow approach velocity is compensated by the
increase in the thermal De-Broglie wavelength.

There has been an ongoing interest in the meaning of the third law of thermodynamics
\cite{Blau96,Lansberg97,rubia98,Rose99,chen88,Lansberg89,oppenheim89}. 
The issue at stake has been: is the third law an independent 
postulate or it is a consequence of the second 
law and the vanishing of the heat capacity.
This study presents a dynamical interpretation of the third law.
The absolute temperature cannot be reached because the maximum rate 
of cooling vanishes linearly at least with temperature.

\acknowledgements

This research was supported by  the  US Navy under contract 
number N00014-91-J-1498.
The authors want to thank Jeff Gordon for his continuous help, 
discussions and  willingness to clarify many fine points.
T.F. thanks Sylvio May for his help. 
\pagebreak

\appendix

\section{Analysis for the 'moving' Cycles.}
\label{sec:MovingCycles}

\begin{figure}[tb]
\vspace{-.7cm}
\psfig{figure=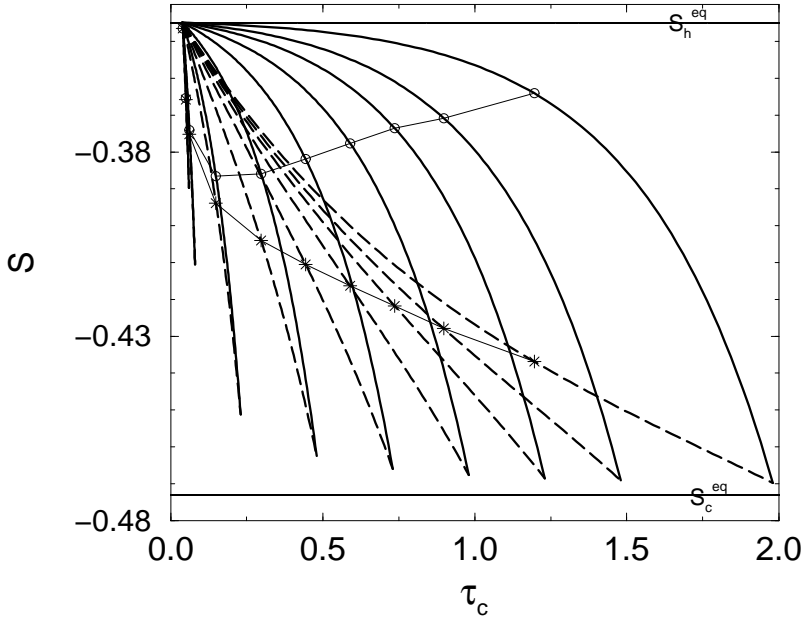,width=0.7\textwidth}
\vspace{0.1cm}
\caption{Comparison between the polarizations S$_1$ and S$_2$
as a function of $\tau_c$, for ten different $ \tau$
values, 0.06, 0.08, 0.1, 0.25, 0.5, 0.75, 1 1.25, 1.5 and 2. 
The solid curves are S$_1$ while the dashed
curves are  S$_2$. Superimposed  are the values  of  S$_1$
and   S$_2$ for the maximal  S$_1$~-~ S$_2$.     }
\label{fig:s1s2t1}
\end{figure}

Insight into the origin of the behavior of the 'moving' cycles is seen in 
Fig. \ref{fig:s1s2t1}, where the polarizations 
S$_1$, S$_2$ are shown as monotonically
decreasing functions of the time allocation
on the cold bath. However, the envelope of S$_1$ for maximal
power, namely for maximal S$_1$-S$_2$ is
worth noticing. It is a decreasing
function for short cycle times, achieves a minimum at  $ \tau_{0}$,
and starts to increase for $ \tau~>~ \tau_{0} $.
Thus it is responsible for  shifting
the cycles to smaller polarization for short cycle times, and for the change
of that trend for larger cycle times.
The envelope of S$_2$ for maximal S$_1$-S$_2$ is also a
monotonically decreasing function of $ \tau_c$, or equivalently of
$ \tau$, supporting the increase of  S$_1$-S$_2$.  
The figure also shows, that for a short time allocation
both S$_1$ and S$_2$ are close to the equilibrium polarization 
S$_h^{eq}$, When not enough time is allocated on the hot bath  both the
polarizations S$_1$ and S$_2$ approach S$_c^{eq}$.

\section{The computation of $ \tau_0$}
\label{sec:Apptau0}

The computation of $ \tau_0$
Eq.  (\ref{WcycH11}) is not sufficient since  it gives only
the relation between the times spent on the cold and hot
branches for zero work. The natural additional
requirement is to seek for the optimal allocations,
$ \tau_{c,0}$ and $ \tau_{h,0}$ using Eq.  (\ref{optimal10f}): 
$ \tau_{0}$~=~$ \tau_{c,0}$~+~$ \tau_{h,0}$~+~$ \tau_a$~+~$ \tau_b$

Denoting by $x_0$ and $y_0$ the corresponding x and y  values
defined in Eq.  (\ref{xy}), the following
two equations for  x$_0$ and y$_0$ are obtained:
\begin{eqnarray}
y_{0} ~=~{  ( x_{max}~-~x_{0})  ~-~  
        { R    }  
\over 
  ( x_{max} ~-x_{0})  ~-~  
        { Rx_{0} } }
\label{opty0}
\end{eqnarray}
and
\begin{eqnarray}
\Gamma_c ~x_{0} \left( (1-y_{0}) (1~-~y_{0}~x_{max}) \right)
~=~
\Gamma_h ~y_{0} \left( (1-x_{0}) (x_{max}~-~x_{0}) \right)
\label{optimal0f}
\end{eqnarray}  
Where R is defined as:
\begin{eqnarray}
R ~=~{   \sigma^2~\omega_a~(1/\tau_a~+~1/\tau_b)~  
\over 
(\omega_b~-~\omega_a)~( S_h^{eq} ~-S_c^{eq}  ~+~  
      \sigma^2/{\tau_a} ) }
\label{opty0b}
\end{eqnarray}
and x$_{max}$ was defined in Eq.   (\ref{minxc}) as:
\begin{eqnarray}
x_{max} ~=~{  ( S_h^{eq}~-~S_c^{eq})  ~-~  
        { \sigma^2/\tau_b    }  
\over 
  ( S_h^{eq} ~-S_c^{eq})  ~+~  
         { \sigma^2/\tau_a } }
\label{xmaxo}
\end{eqnarray}
The quadratic equation to be solved for x$_0$ is,
\begin{eqnarray}
AA~x_{0}^2~+~BB~x_{0}~+~CC~=~~0
\label{optx00}
\end{eqnarray}  
Where  AA~=~$\Gamma_h$~(1~+~R)~~
       BB~=~-~($\Gamma_h$~((1~+~R)~(x$_{max}$-R)~+~x$_{max}$)
              ~+~$\Gamma_c$~(1~+~R~-~$x_{max}$))
and   CC~=~$\Gamma_h$~(x$_{max}$-R)~x$_{max}$

\section{Entropy production.}
\label{sec:entrfricA}

(1) \bf Heat Engine. \rm

\begin{eqnarray}
\Delta {\cal S}^u_{cyle1}~=~ -({\cal Q}_{AB}/T_h~+~{\cal Q}_{CD}/T_c)
\label{DScycH1}
\end{eqnarray}
Or from Table (\ref{tab:cycle1})
\begin{eqnarray}
\Delta {\cal S}^u_{cyle1}
~=~(\omega_a/T_c-\omega_b/T_h) (S_1-S_2)
~+~{ \sigma^2 \omega_a  \over {T_c} } (1/\tau_a~+~1/\tau_b)
\label{DScycH2}
\end{eqnarray}
  
\begin{figure}[tb]
\vspace{-.2cm}
\psfig{figure=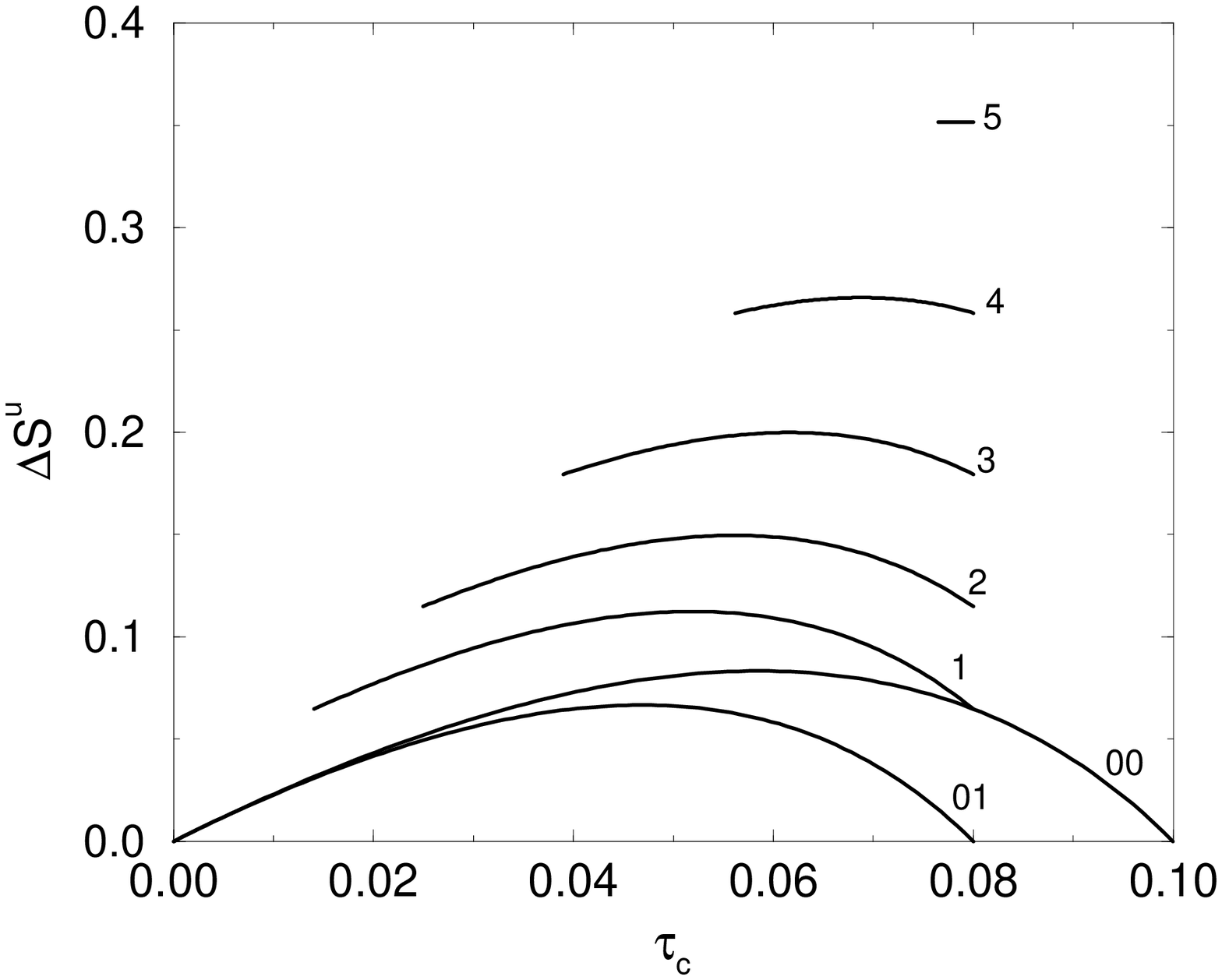,width=0.3\textwidth}
\vspace{-4.1cm}
\hspace{4.8cm}
\psfig{figure=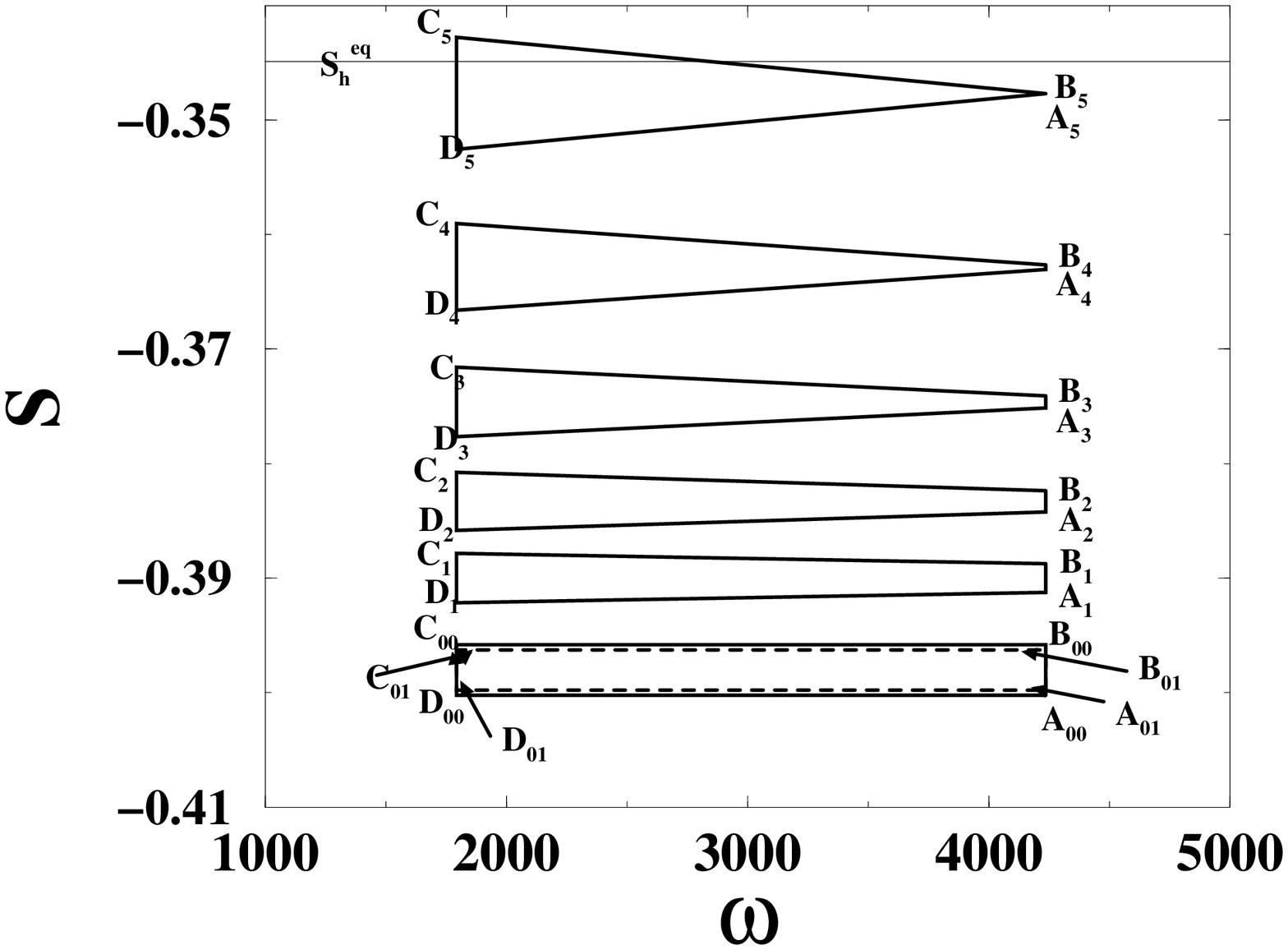,width=0.3\textwidth}
\vspace{-4.1cm}
\hspace{.1cm}
\psfig{figure=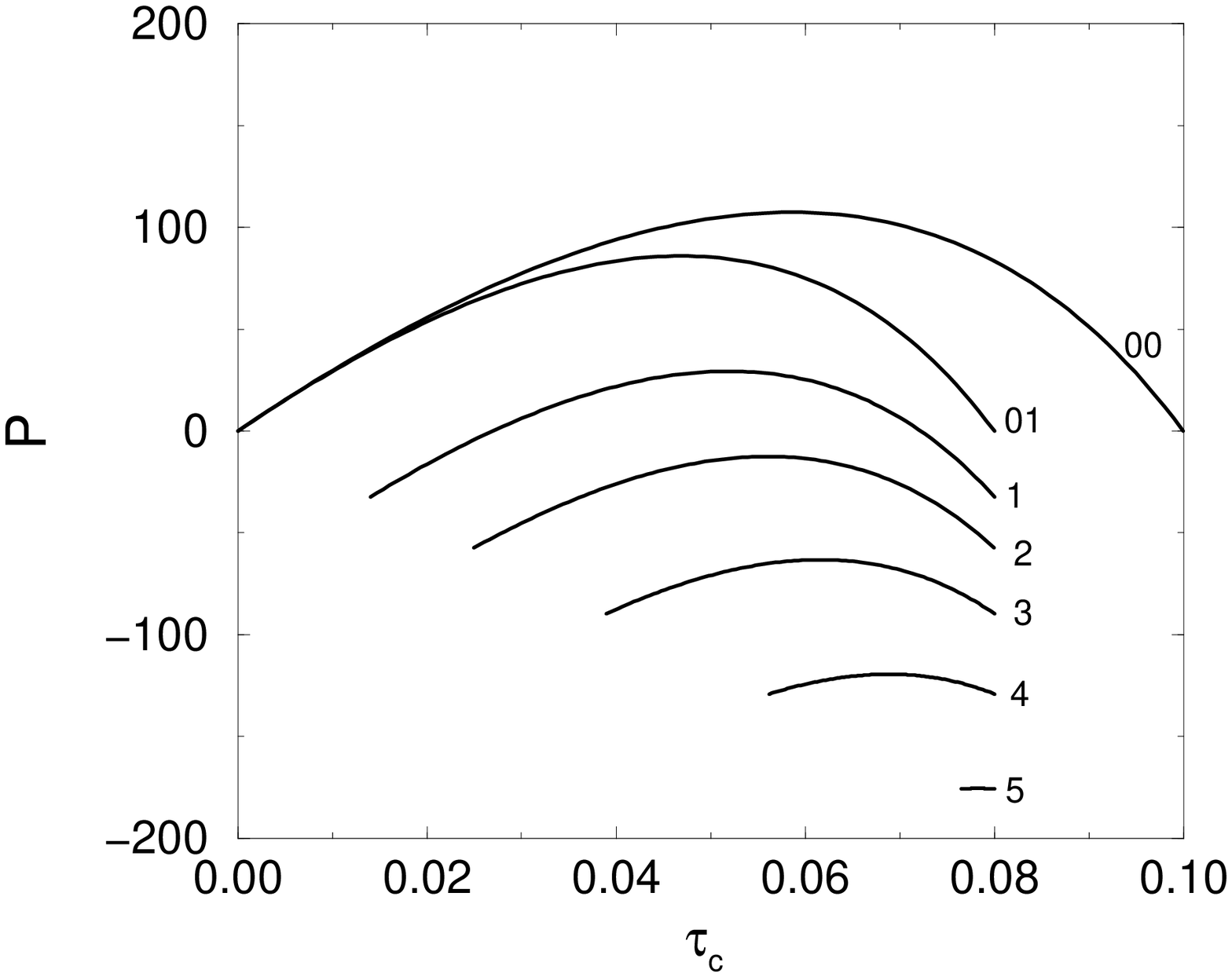,width=0.3\textwidth}
\vspace{4.4cm}
\caption{Left: The entropy production of the heat engine as a function 
of the time spent on the cold branch for the fixed values of
$\omega_a=1794$, $\omega_b=4238$, $T_c=500$, $T_h=2500$,
$\Gamma_c=1$,~$\Gamma_h=2$~
and $\tau=0.1$.
Middle: The corresponding cycles.~Right: The corresponding powers.
Seven cases are shown. Case~'00' is the frictionless case, when the 
times spent on the 'adiabats', are zero, case~'01' 
 is the frictionless case, when the 
times spent on the 'adiabats',  $\tau_a$ and $\tau_b$  
are different from zero and equal: 0.01. The other five cases are with
increasing friction, when also  $\tau_a = \tau_b~=~0.01 $, whereas the 
different friction coefficients $\sigma$ are: for
 plot~ 1: $\sigma=0.003$, for~plot~ 2: $\sigma=0.004$,~
for plot~ 3: $\sigma=0.005$,
for~plot~4: $\sigma=0.006$ and for~plot~ 5: $\sigma=0.007$.    }
\label{fig:entr1}
\end{figure}

The entropy production results are shown in Fig.  \ref{fig:entr1}.  
The left figure shows   $ \Delta {\cal S}^u$ 
with increasing friction. The middle figure shows the corresponding cycles,
while the right figure shows the corresponding power values.

The reciprocal behavior of the entropy production and the power
is clear from Fig. \ref{fig:entr1}. One also observes, 
that for the given cycle time the 'free' time for the cycles 
with increasing $ \sigma$ becomes more restricted. This
follows from the dependence of $ \tau_{c,min}$ 
on $ \sigma$. See also  Fig.( \ref{fig:taumin7})

Introducing Eq. (\ref{eqS1S2}) into  Eq. (\ref{DScycH2}).
The entropy production becomes,
\begin{eqnarray}
\Delta {\cal S}^u_{cyle1}~=~(\omega_a/T_c-\omega_b/T_h) 
(S_h^{eq}-S_c^{eq})F(x,y)~+~\Delta {\cal S}^u_{\sigma1}
\label{DScycH3}
\end{eqnarray}
where
\begin{eqnarray}
\Delta {\cal S}^u_{\sigma1}~=~  \sigma^2 {1 \over (1-xy) } \left( 
{ { \omega_a \over T_c} } { (1-x) } (1/\tau_a~+~y/\tau_b)~+~
{ { \omega_b \over T_h} } { (1-y) } (x/\tau_a~+~1/\tau_b) \right)
\label{DScycH4}
\end{eqnarray}
Notice, that $\Delta {\cal S}^u_{\sigma1}$ is always positive.
For $\sigma = 0$ Eq. (\ref{DScycH4}) reduces to the 
frictionless results  \cite{feldmann96}.

(2) \bf Heat Pump  \rm

The entropy production for the heat pump becomes:
\begin{eqnarray}
\nonumber
{\Delta {\cal S}^u}_{ref} = \left( { \omega_b \over T_h } - 
{ \omega_a \over T_c }  \right)
 (S_2-S_1)~+~\sigma^2
{ \omega_b~ \over T_h }  (1/\tau_a~+~1/\tau_b)\\
~~=~~\left( { \omega_b \over T_h }-{ \omega_a \over T_c }  \right)
(S_2^{eq}-S_1^{eq})\cdot F(x,y)
\nonumber
~+~\\
\sigma^2 F(x,y)\left\{\frac{\omega_b}{T_h}\frac{1}{1-x}
(\frac{1}{\tau_a}~+~ \frac{x}{\tau_b})
~+~
\frac{\omega_a}{T_c}\frac{1}{1-y}
(\frac{1}{\tau_b}~+~ \frac{y}{\tau_a} ) \right\}
\label{entref}
\end{eqnarray}

The asymptotic entropy production 
as T$_c$ tends to zero can be calculated leading to 
\begin{eqnarray}
{\Delta {\cal S}^u_{ref}}~=~F(x,y)~[
(\omega_b/(\rho \omega_a)) 
(1-\rho (\omega_a/\omega_b))^2
\nonumber
~+~\\ 
\sigma^2
~\left( { \omega_b \over \rho T_c }
{1 \over(1-x) } (1/\tau_a+x/\tau_b)~+~
{ \omega_a~ \over T_c } 
{1 \over(1-y) } (1/\tau_b+y/\tau_a)
  \right)]
\label{entrefas}
\end{eqnarray}

Since  $T_h=~\rho T_c$, the r.h.s. of Eq. (\ref{entrefas}) 
tends to a constant, for
each term depends on the constant ratios  
($\omega_b$/T$_h$), ($\omega_a$/T$_c$) or on their ratio. 
This result is demonstrated on  the right side of Fig. \ref{fig:asympent1}.

The optimization with respect to time
allocation has  the same result as for the heat engine. Therefore,
only optimization with respect to the fields are presented;

Equating to zero the derivatives with respect to x an y
of the entropy production, one gets two  similar equation to
the total work derivatives: 
\begin{eqnarray}
~{ (1~-~y~x_{max}) \over ({\omega_a}/{T_c}~-~{\omega_b}{T_h})}
~({ \Delta S^{eq}~+~ \sigma^2/\tau_a})
~  \cosh^2 \left( {{\omega_a} \over {2 k_B T_c} } \right)~+~
{~1-y \over (4~k_B~T_c) } ~\geq 0
\label{optimale6}
\end{eqnarray}

\begin{eqnarray}
~{ (x_{max}~-~x) \over ({\omega_a}/{T_c}~-~{\omega_b}{T_h})}
~({ \Delta S^{eq}~+~ \sigma^2/\tau_a})
~  \cosh^2 \left( {{\omega_b} \over {2 k_B T_h} } \right)~+~
{~1-x \over (4~k_B~T_h) } ~\geq 0
\label{optimale7}
\end{eqnarray}

Where $ \Delta S^{eq}$, is  $S_h^{eq}$- $S_c^{eq}$.
 
Eqs.  (\ref{optimale6}) and (\ref{optimale7}) show
that the entropy production is a monotonic function in the allowed
range, namely, for
\begin{eqnarray}
{ \omega_a \over T_c } > { \omega_b \over T_h }. 
\label{optimal6}
\end{eqnarray}   
                                                             
To conclude the entropy production has a minimum value: 
$\Delta {\cal S}^u_{min}$, will be 
\begin{eqnarray}
{\Delta {\cal S}^u_{min}}~=~ { \omega_a \over T_c } \sigma^2 (1/\tau_a+1 / \tau_b)
\label{mindsu}
\end{eqnarray} 
obtained on the boundary of the range.

\section{The Total Work done on the System for the Heat Pump}
\label{totalw}

The total work done on the system becomes,

\begin{eqnarray}
{\cal W}_{cyle3}^{on} = (\omega_b-\omega_a) (S_2-S_1)~+~\sigma^2
\omega_b~ (1/\tau_a~+~1/\tau_b)
\label{WcycRf0}
\end{eqnarray}
or
\begin{eqnarray}
{\cal W}_{cyle3}^{on} = (\omega_b-\omega_a) (S_2^{eq}-S_1^{eq}) F(x,y)~+~
W_{\sigma3}
\label{WcycRf1}
\end{eqnarray}
where
\begin{eqnarray}
{\cal W}_{ \sigma3 }~=~  { \sigma^2 \over (1-xy)} \left(
\omega_b (1-y)( {1 /\tau_a}~+~{ x /\tau_b} )~~+~~
\omega_a (1-x)( {y /\tau_a}~+~{ 1 / \tau_b} )
\right)~~
\nonumber
\\
~~=~~\sigma^2 F(x,y)\left\{\frac{\omega_b}{1-x}
(\frac{1}{\tau_a}~+~ \frac{x}{\tau_b})
~+~
\frac{\omega_a}{1-y}
(\frac{1}{\tau_b}~+~ \frac{y}{\tau_a} ) \right\}
\label{WcycRf}
\end{eqnarray}
Eq. (\ref{WcycRf0}) can be interpreted  as the work done on the
working fluid  see 
(Cf. Fig. \ref{fig:cycle2}), as
the sum of three positive areas,
$ ( \omega_b~-~\omega_a) (S_2~-~S_1)$, 
$ \sigma^2 \omega_b (1/\tau_a)$~and~ $ \sigma^2 \omega_b (1/\tau_b)$
with the corresponding corners, D,C,B$^1$,A$^1$,~~B,B$^1$,S$_2$,S$_3$
and A$^1$,A,S$_4$,S$_1$.
  
\section{The optimal cooling strategy close to the absolute zero temperature}
\label{strategy}

The first step in  the cooling strategy is to create the first optimal
quartet;
\begin{itemize}

\item{(0) The systems external  parameters $\sigma$, $\tau_a$, $\tau_b$,
$\Gamma_c$ and $\Gamma_h$ are set.}

\item{(1) A decreasing set of $\omega_b$ is chosen.}
 
\item{(2) A constant ratio ($\rho$) for T$_h$/T$_c$, is chosen
which is the ratio of the initial bath temperatures.}

\item{(3) For the above chosen values, the optimal values of
$\omega_a$, T$_c$, $\tau$, and its optimal allocations between the
branches to give maximal Q$_F$ are found
for each $\omega_b$ in the set in (1) }, by solving numerically the
following additional equation to   Eq. (\ref{rheatf6}),
with the condition that  T$_h~=~\rho \cdot$T$_c$:

\begin{eqnarray}
~{ {\partial {{\cal Q}_F}} \over { {\partial {T_c} } } }~=
~{F(x,y)~ \over \ {4~ \tau~ ~k_B}~}~\left(~{ \omega_a \over
T_c} \right)^2 
~\left(
 { 1 \over 
\cosh^2{ \omega_a
\over 2~k_B~T_c } }~-~
 { \omega_b  \over \rho \omega_a~
\cosh^2{ \omega_b
\over 2\rho~k_B~T_c } } \right)~=~0 
\label{rheatf27}
\end{eqnarray}
\end{itemize}

\begin{figure}[tb]
\vspace{-.2cm}
\psfig{figure=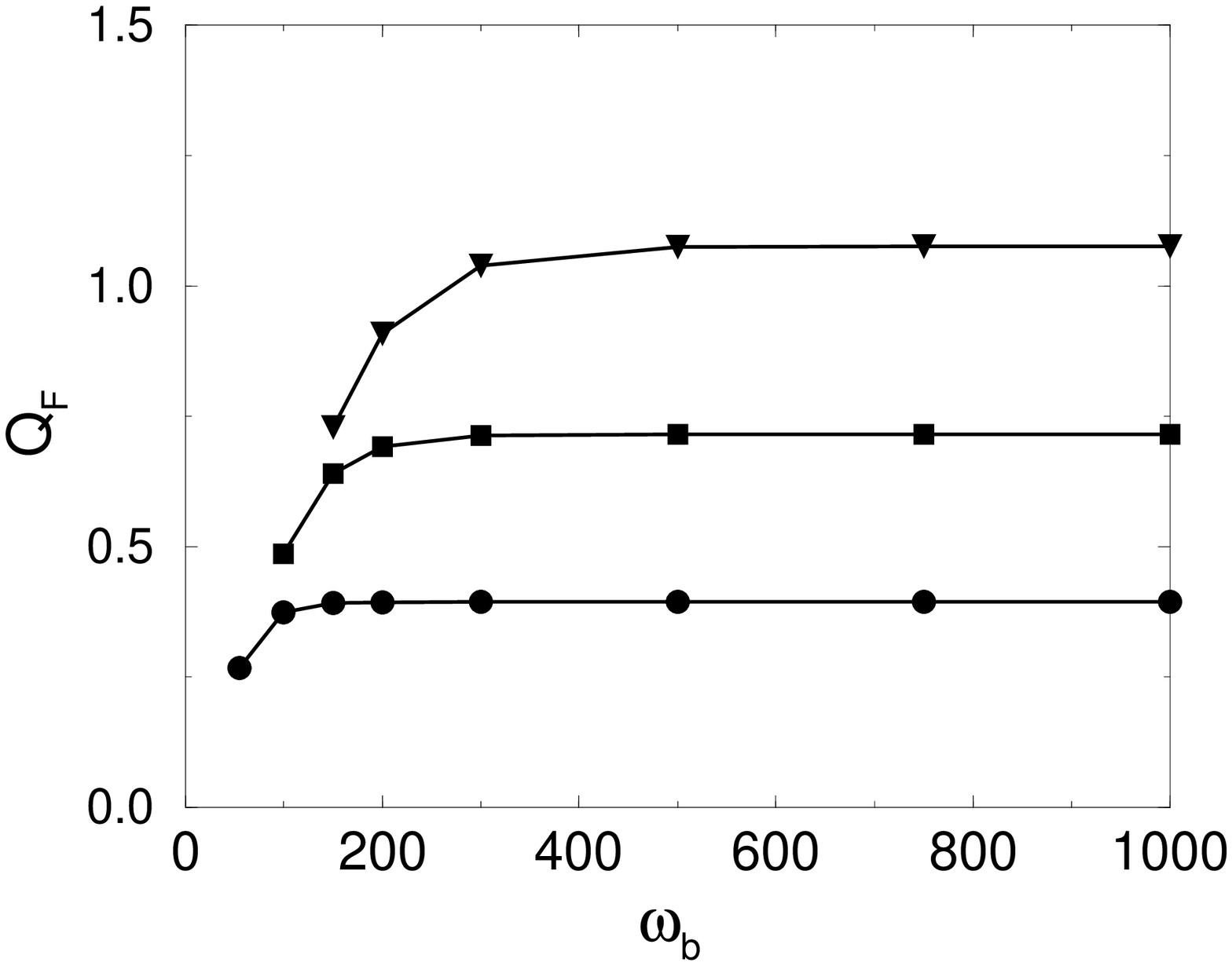,width=0.5\textwidth}
\vspace{-6.78cm}
\hspace{7.cm}
\psfig{figure=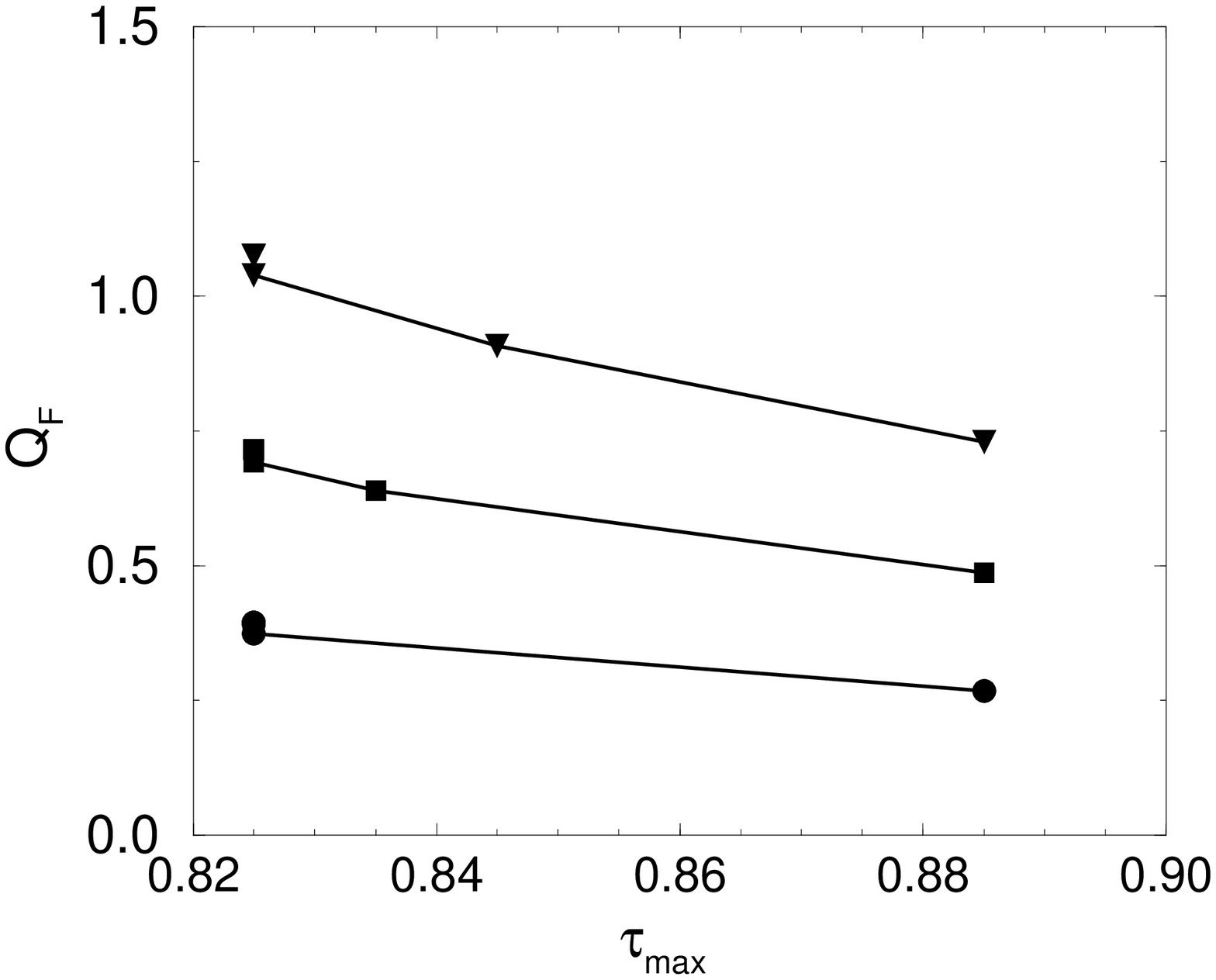,width=0.5\textwidth}
\vspace{0.1cm}
\caption{Left: The optimal heat-flow for the heat pump  
as a function
of $\omega_b$, showing the saturation phenomenon.
The fixed parameter values are: for triangles;
$T_h=64.5725$, $T_c=12.9145$, $\omega_a=11.9233$,~
starting with  $\omega_b~=~150$,
for squares;
$T_h=42.90815$, $T_c=8.58168$, $\omega_a=7.94986$,~
starting with  $\omega_b~=~100$, for circles;
$T_h=23.59905$, $T_c=4.71981$, $\omega_a=4.37247$,~
starting with  $\omega_b~=~55$.
The common parameters for all three figures are:
$ \tau_a = \tau_b =0.01$, $ \sigma=0.005$, $ \Gamma_c=1$, 
$ \Gamma_h=2$. 
Right: The optimal heat-flows  as a function of $ \tau_{max}$,
the time at which the optimum is achieved. 
The fixed parameter values are the same, as on the left.
We note, that the optimal time is becoming constant 
only at saturation.   }
\label{fig:qftcwamax}
\end{figure}

The above strategy causes the decrease of $T_h$ together with 
the $T_c$. Nevertheless according to (ii) above, the doublet 
$\omega_b$ and $T_h$ can be rescaled to increase $T_h$
back to its original value.
The solid curves of Fig.  \ref{fig:qftcwa1} are optimal in the 
in the above described sense.
Increasing {\em only} the value of $\omega_b$ in the optimal
quartet according to point (iii), leads to larger values of 
the cooling rate,
but eventually the increase of ${\cal Q}_F$ will slow down and saturate.
See Fig.   \ref{fig:qftcwamax}  and the dashed curves of
Fig.  \ref{fig:qftcwa1}.

Fig.  \ref{fig:qftcwamax}
represents the saturation phenomenon on $\omega_b$.  
Three points from 
Fig.  \ref{fig:qftcwa1} are chosen, and all parameters
are fixed, except $\omega_b$,
which is allowed to increase.
\newpage 

In order to approach the upper-bound for ${\cal Q}_F$
in Eq. (\ref{rheatf22}), 
a decreasing set of $\omega_a~/~T_c$ is created, 
achieved in an optimal way:

First step: After having  an optimal 'quartet',
T$_c$ and T$_h$, are fixed. Then, by lowering $\omega_b$,
one finds the corresponding optimal   $\omega_a$ values.
This procedure is checked
globally, by also iterating the time allocations.
The results of a typical example are shown in  
Table \ref{tab:zeroset}.

Second step: 
Using again the property of extensivity, the cooling will be 
achieved by multiplying the rows of Table \ref{tab:zeroset}
by a decreasing sequence,
e.g. by $2^{-n}$ for the n-th row. Table \ref{tab:zeroset2}
describes  the cooling strategy, checking also the
non-divergence of the entropy production both for the frictionless
case and the case with friction.
The results are also summarized in  Fig. \ref{fig:asympent1}.

Table \ref{tab:zeroset}
demonstrates, that the procedure shifts down to the Carnot
bound. 
The ratio R~=~ $ {  \omega_b \over T_h } $/ $ {  \omega_a \over T_c } $.
was computed showing  only small changes.

\begin{table}
\caption{First step. Starting from an optimal quartet, the procedure 
creates for a given decreasing set of $ \omega_b$-s, a 
decreasing  set of of $\omega_a$-s for fixed bath temperatures. }    
\begin{center}
\begin{tabular}{||c|c|c|c|c|c|c|c||}
\tableline
$T_c $ & T$_h$ & $ \omega_b$ &  $\omega_a^{optimal}$ &     
 $\omega_a^{optimal}/T_c$ & $ \omega_b/T_h$ & R & ${\cal Q}_F $  \\
\tableline
0.0025  & 50 & 60  & 1.370(-3) & 0.5392  & 1.2 & 2.226 & 5.812(-5)  \\
\tableline
0.0025  & 50 & 55  & 1.273(-3) & 0.5090  &  1.1  & 2.161 & 5.013(-5)  \\
\tableline
0.0025  & 50 & 50  &  1.164(-3) & 0.4653   & 1 & 2.149 &  4.241(-5)  \\
\tableline
0.0025  & 50 & 45 & 1.051(-3) & 0.4205 & 0.9 & 2.140    &  3.50549(-5)  \\
\tableline
0.0025  & 50 & 40 & 9.320(-4) & 0.3728  & 0.8 & 2.146    & 2.826(-5)  \\
\tableline
0.0025  & 50 & 35 & 8.250(-4) & 0.3300  & 0.7 & 2.121 & 2.182(-5)  \\ 
\tableline
0.0025  & 50 & 30 & 6.985(-4) & 0.2794 & 0.6 & 2.147 & 1.613(-5)  \\ 
\tableline
\end{tabular}
\end{center}
\label{tab:zeroset}
\end{table}

\begin{table}
\caption{A procedure to get an optimal set of pairs 
of $\omega_a,~T_c$
where their ratio tends to zero. T$_h =~50$ for every cold bath temperature,
T$_c$. The index 'fl' stands for the frictionless case, and  $Q_F^{up}$
denotes the upper-bound for $ {\cal Q}_F$.  }    
\begin{center}
\begin{tabular}{||c|c|c|c|c|c|c||}
\tableline
$T_c $ & $ \omega_b$ &        
$\omega_a^{optimal}$ & $\Delta S^u$ & $\Delta S^{u,fl}$ & ${\cal Q}_F$ & ${\cal Q}_F^{up}$  \\
\tableline
0.0025  & 60 & 1.370(-3) & 0.0333353 & 0.0285 & 5.812(-5) & 6.084(-5)  \\
\tableline
0.00125 & 55 & 6.365(-4) & 0.0285075 & 0.02406 & 2.457(-5) & 2.626(-5)  \\
\tableline
0.000625 & 50 & 2.91(-4) & 0.0242662 & 0.02021 & 1.039(-5) & 1.098(-5)  \\
\tableline
0.0003125 & 45 & 1.3138(-4) & 0.020338 & 0.01669 & 4.2939(-6) & 4.467(-6)  \\
\tableline
0.00015625 & 40 & 5.825(-5) & 0.016788 & 0.01354 & 1.7239(-6) & 1.759(-6)  \\
\tableline
0.0000781  & 35 & 2.578(-5) & 0.013210 & 0.010357 & 6.6772(-7) & 6.888(-7)  \\ 
\tableline
0.0000391  & 30 & 1.0914(-5) & 0.010356 & 0.007915 & 2.4673(-7) & 2.468(-7) \\
\tableline
\end{tabular}
\end{center}
\label{tab:zeroset2}
\end{table}


\begin{references}

\bibitem{carnot}
S. Carnot, {\em R\'2 sur la Puissance Motrice du Feu et sur les
  Machines propres \`{a} D\'{e}velopper cette Puissance} 
(Bachelier, Paris, 1824).

\bibitem{curzon75}
F.L. Curzon and B. Ahlborn, Am. J. Phys. {\bf 43}, 22 (1975)

\bibitem{salamon77}
P.Salamon, B. Andresen and R.S. Berry
Phys. Rev. A {\bf 15}, 2094 (1977)

\bibitem{salamon80}
P. Salamon, A. Nitzan, B. Andresen and R.S. Berry
Phys. Rev. A {\bf 21}, 2115 (1980)

\bibitem{andresen83}
B. Andresen, "Finite-Time Thermodynamics", 
(Phys. Lab II. University of Copenhagen, Copenhagen 1983).

\bibitem{bejan96}
A. Bejan, "Entropy Generation Minimization", 
(Chemical Rubber Corp., Boca Raton FL. 1996).

\bibitem{szilard29}
L. Szilard,  Z. Physik {\bf 53}, 840 (1929). 

\bibitem{brillouin}
{Leon Brillouin}, 
\newblock {"Science and Information Theory"},
\newblock {Academic Press,  1956}. 

\bibitem{lloyd}
S. Lloyd, Phys. Rev. A {\bf 56} 3374 (1997).


\bibitem{geusic59}
J. Geusic, E.~S. du~Bois, R.~D. Grasse, and H. Scovil, 
J. App. Phys. {\bf 30},  1113  (1959).

\bibitem{scovil59}
H. Scovil and E.~S. du~Bois, Phys. Rev. Lett. {\bf 2},  262  (1959).


\bibitem{geusic67}
J. Geusic, E.~S. du~Bois, and H. Scovil, Phys. Rev. {\bf 156},  343  (1967).


\bibitem{levine74}
R. D. Levine and O. Kafri, Chem. Phys. Lett. {\bf 27},  175  (1974).

\bibitem{benshaul79}
A. Ben-Shaul and R.D. Levine, J. Non-Equilib. Thermodyn. {\bf 4},  363  (1979).

\bibitem{k24}
{R. Kosloff},
\newblock {A Quantum Mechanical Open System as a Model of a Heat Engine.},
\newblock J. Chem. Phys., {\bf 80},\hspace{0.25em}1625  (1984).

\bibitem{geva0}
{E. Geva and R. Kosloff},
\newblock {"A Quantum Mechanical Heat Engine Operating in Finite Time. A Model
  Consisting of Spin $~half~$ Systems as The Working Fluid"},
\newblock J. Chem. Phys., {\bf 96},\hspace{0.25em}3054  (1992).

\bibitem{geva1}
{E. Geva and R. Kosloff},
\newblock {"On the Classical Limit of Quantum Thermodynamics in Finite Time"},
\newblock J. Chem. Phys., {\bf 97},\hspace{0.25em}4398  (1992).

\bibitem{geva2}
{E. Geva and R. Kosloff},
\newblock {The Quantum Heat Engine and Heat Pump: An Irreversible Thermodynamic
  Analysis of The Three-Level Amplifier},
\newblock J. Chem. Phys., {\bf 104},\hspace{0.25em}7681  (1996).

\bibitem{wu98}
{F. Wu, L. G. Chen, F. R. Sun, C. Wu and P. Q. Hua},
\newblock {"Optimal performance parameters for a quantum Carnot heat pump with spin-1/2"},
\newblock Energy Conversion and Management, {\bf 39},\hspace{0.25em}1161  (1998).

\bibitem{feldmann96}
T. Feldmann, E. Geva, R. Kosloff and P. Salamon, 
"Heat engines in finite time governed by master equations",
Am. J. Phys. {\bf 64}, 485 (1996)
  


\bibitem{Velasco}
S. Velasco J. M. M. Roco, A. Medina and A Calvo Hernandez,
"New Performance Bounds for a Finite-Time Carnot Refrigerator"
Phys. Rev. Lett. {\bf 78} 3241 (1997).



\bibitem{Velasco97}
S. Velasco J. M. M. Roco, A. Medina and A Calvo Hernandez,
"Irreversible refrigerator under per-unit-time coefficient 
of performance optimization"
Appl. Phys. Lett.  {\bf 71}, 1130 (1997).

\bibitem{Velasco98}
A. Calvo Hernandez J. M. M. Roco, S. Velasco, A. Medina and ,
"Irreversible Carnot cycle under per-unit-time efficiency optimization."
Appl. Phys. Lett.  {\bf 73}, 853 (1998).

\bibitem{yan98}
Z. Yan and J. Chen,
"Comment on "New Performance Bounds for a Finite-Time Carnot Refrigerator",
Phys. Rev. Lett. {\bf 81} 5469 (1998).


\bibitem{chen98}
J. C. Chen and Z. J. Yan,
"The effect of thermal resistances and regenerative losses on the 
performance characteristics of a magnetic Ericsson refrigerator cycle.",
J. Appl. Phys.  {\bf 84} 1791 (1998).

\bibitem{gordon91}
J. M. Gordon and M. Huleihil,
"On optimizing maximum-power heat engines"
J. Appl. Phys. {\bf 69}, 1 (1991).

\bibitem{gordon97}
J. M. Gordon, K. C. Ng and H.T. Chua, 
"Optimizing chiller operation based on finite-time
thermodynamics: universal modeling and experimental confirmation",
Int. J. Refrg. {\bf 20}, 191 (1997).

\bibitem{cohen}
C. Cohen-Tanoudji,
"Manipulating atoms with photons",
Rev. Mod. Phys. {\bf 70}, 707 (1998).

\bibitem{forrey}
N. Balakrishnan, R. C. Forrey, and A. Dalgarno,
"Quenching of H$_2$ Vibrations in Ultracold ~$^3$He and ~$^4$He Collisions",
Phys. Rev. Lett. {\bf 80}, 3224 (1998).

\bibitem{Blau96}
S. Blau and B. Halfpap, "Question no 34, "What is 
the Third law of thermodynamics trying to tell us?"
Am. J. Phys. {\bf 64}, 13 (1996).

\bibitem{Lansberg97}
P. T. Landsberg , "Answer to Question no 34, "What is 
the Third law of thermodynamics trying to tell us?"
Am. J. Phys. {\bf 65}, 269 (1997).

\bibitem{rubia98}
S. Mafe and J. De la Rubia, , "Answer to Question no 34, "What is 
the Third law of thermodynamics trying to tell us?"
Am. J. Phys. {\bf 66}, 277 (1998).

\bibitem{Rose99}
C. Rose-Innes, , "Answer to Question no 34, "What is 
the Third law of thermodynamics trying to tell us?"
Am. J. Phys. {\bf 67}, 273 (1999).

\bibitem{chen88}
Z. J. Yan and J. C. Chen,
"An equivalent theorem of the Nernsts Theorem"
J. Phys. A  {\bf21} L707 (1988).

\bibitem{Lansberg89}
P. T. Landsberg, 
"A comment on Nernsts Theorem"
J. Phys. A  {\bf22} 139 (1989).

\bibitem{oppenheim89}
I. Oppenheim
"A comment on an equivalent theorem of the Nernsts Theorem"
J. Phys. A  {\bf22} 143 (1989).





\end{references}
\end{document}